\documentclass[a4paper,fleqn]{cas-dc}
\usepackage{amsmath} % for math formatting
\usepackage[numbers,sort&compress]{natbib}
\usepackage[font=small,skip=0pt]{caption}
\usepackage{array}
\usepackage{hyperref}
\usepackage{float}
\usepackage{bm}
\usepackage{algorithm}
\usepackage{algpseudocode}
\usepackage{subcaption}
\def\tsc#1{\csdef{#1}{\textsc{\lowercase{#1}}\xspace}}
\tsc{WGM}
\tsc{QE}
\begin{document}
\let\WriteBookmarks\relax
\def\floatpagepagefraction{1}
\def\textpagefraction{.001}

% Short title
\shorttitle{Joint mode switching and resource allocation in wireless-powered RIS-aided multiuser communication systems}    

% Short author
\shortauthors{Mingang Yuan \emph{et al.}}   

% Main title of the paper
\title [mode = title]{Joint mode switching and resource allocation in wireless-powered RIS-aided multiuser communication systems}  

% Title footnote mark
% eg: \tnotemark[1]
%\tnotemark[<tnote number>] 

% Title footnote 1.
% eg: \tnotetext[1]{Title footnote text}
%\tnotetext[<tnote number>]{<tnote text>} 

% First author
%
% Options: Use if required
% eg: \author[1,3]{Author Name}[type=editor,
%       style=chinese,
%       auid=000,
%       bioid=1,
%       prefix=Sir,
%       orcid=0000-0000-0000-0000,
%       facebook=<facebook id>,
%       twitter=<twitter id>,
%       linkedin=<linkedin id>,
%       gplus=<gplus id>]

\author[1]{Mingang Yuan}[orcid=0009-0003-9561-1612]
\ead{2112230009@e.gzhu.edu.cn}

\author[1]{Wenzhe Zhang}
\ead{2112330009@e.gzhu.edu.cn}

\author[1]{Gaofei Huang}
\ead{huanggaofei@gzhu.edu.cn}
\cormark[1]

% Footnote of the first author
%\fnmark[<footnote mark no>]

% Email id of the first author
%\ead{<email address>}

% URL of the first author
%\ead[url]{<URL>}

% Credit authorship
% eg: \credit{Conceptualization of this study, Methodology, Software}
%\credit{<Credit authorship details>}

% Address/affiliation
\affiliation[1]{organization={School of Electronics and Communication Engineering, Guangzhou University},
city={Guangzhou}, 
postcode={510006}, 
country={China}}

%\author[<aff no>]{<author name>}[<options>]

% Footnote of the second author
%\fnmark[2]

% Email id of the second author
%\ead{}

% URL of the second author
%\ead[url]{}

% Credit authorship
%\credit{}

% Address/affiliation
%\affiliation[<aff no>]{organization={},
%            addressline={}, 
%            city={},
%          citysep={}, % Uncomment if no comma needed between city and postcode
%            postcode={}, 
%            state={},
%            country={}}

% Corresponding author text
\cortext[1]{Corresponding author:Gaofei Huang}

% Footnote text
%\fntext[1]{}

% For a title note without a number/mark
%\nonumnote{}

% Here goes the abstract
\begin{abstract}
This paper investigates a wireless-powered hybrid reflecting intelligent surface (hybrid RIS)-assisted multiple access system, where the RIS can harvest energy from energy station (ES) transmitted radio frequency signal (RF), and each reflecting element can flexibly switch between active mode, passive mode, and idle mode. The objective is to minimize the maximum energy consumption of the users by jointly optimizing the operating modes of each reflecting element, the amplification factor of active elements, the transmit power, and transmission time allocation, subject to quality-of-service (QoS) of each user and the available energy constraint of RIS. In the formulated optimization problem, the operating modes of each reflecting element are highly coupled with the amplification coefficient of the active reflecting elements, making it a challenging mixed-integer programming problem. To solve this problem, a hierarchical optimization method based on deep reinforcement learning is proposed, where the operating modes of each reflecting element and the amplification coefficient of active elements are obtained by solving the outer sub-problem using proximal policy optimization (PPO), and the transmit power and transmission time allocation are obtained by solving the inner sub-problem using convex optimization methods. Simulation results show that compared to the baseline scheme, the proposed scheme can reduce user energy consumption by $70 \%$.
\end{abstract}

% Use if graphical abstract is present
%\begin{graphicalabstract}
%\includegraphics{}
%\end{graphicalabstract}

% Research highlights
\begin{highlights}
\item
\item 
\item 
\end{highlights}

% Keywords
% Each keyword is separated by \sep

\maketitle
\section{Introduction}
% Main text
\subsection{Background}
Reconfigurable Intelligent Surface (RIS) is an emerging technology that can use metasurfaces of intelligent control to reconfigure poor electromagnetic wave propagation conditions. This technology can effectively enhance the spectral efficiency and energy efficiency of wireless communication systems, making it one of the potential key enabling technologies for 6G \cite{Nine, Tutorial}.  Currently, the vast majority of RIS studied for research are passive, with each reflecting element consisting of passive components such as capacitors and resistors, offering the advantages of low cost and low power consumption. However, the electromagnetic wave signal, reflected by the RIS, must sequentially traverse the transmitter-RIS and RIS-receiver links. This results in the overall path loss of the transmitter-RIS-receiver link being the product of the individual path losses of these two links. Therefore, the signal's power reflected by the passive RIS significantly attenuates when it reaches the receiver. This phenomenon
is known as ‘multiplicative fading’, which greatly limits the performance of communication systems assisted by passive RIS. Scholars have proposed a new type of active RIS architecture to overcome ‘multiplicative fading’ caused by passive RIS \cite{Active_first,Dai}. The active RIS can reflect and amplify the incident signal, effectively addressing the ‘multiplicative fading’ issue attributed to passive RIS.

Nevertheless, active RIS-assisted communication systems face their challenges. The amplification process not only introduces amplification noise but also consumes a lot of energy. According to studies in reference \cite{Same_Power_Budget, Active_Passive, Reflect_or_Not}, when the available energy of active RIS is limited, the system performance may not be optimal and could even be worse than Passive RIS. Based on these considerations, scholars have recently proposed an innovative architecture that integrates active RIS with passive RIS \cite{New_Hybrid,Exploring_Hybrid}, which is known as hybrid active-passive RIS. In this new architecture, each reflecting element can flexibly switch between active and passive modes according to requirements, achieving better system performance than active RIS. 

Furthermore, most existing RIS research assumes that RIS is powered by a stable power source such as batteries or power grid, which limits RIS deployment, particularly in remote or hazardous areas. The integration of RF energy harvesting technology into RIS, which enables them to harvest ambient RF energy for self-sustainability, presents a viable and appealing approach to overcoming these deployment limitations. Therefore, communication systems assisted by wirelessly powered passive \cite{Wireless-Powered,Optimized-Energy,TCCN,Robust-and-Secure,Energy-Harvesting,Computation-Offloading}, active \cite{Performance-Analysis,RIS-Aided-EH-NOMA}, and hybrid \cite{Multi-Functional} RIS have become a hot topic in the field of wireless communications in recent years. 

\subsection{Related Work}
\subsubsection{Research In Active RIS}
Presently, the majority of research in RIS-assisted communication predominantly focused on passive RIS\cite{TVT,UAV,Ruichen}. Concurrently, active RIS-assisted communication has attracted scholarly attention in recent years \cite{Energy_Constrained_IoT,Active_WCL,mao2023energy,Low_Complexity,Active_SWIPT,Wideband,Nearly_Passive}. Reference \cite{Energy_Constrained_IoT} investigated active RIS assisted uplink time division multiple access (TDMA) and non-orthogonal multiple access (NOMA) system, where the sum throughput was maximized by jointly optimizing active RIS beamforming, time allocation and transmit power of device subject to available energy of each device and amplification power of active RIS. References \cite{Active_WCL,mao2023energy} investigated active RIS-assisted wireless-powered communication networks (WPCN) consisting of multiple energy-harvesting terminal users by jointly optimizing time allocation, transmit power of the device and active RIS beamforming maximizing the sum throughput and energy efficient, respectively. Reference \cite{Low_Complexity} proposed low-complexity arithmetic to maximize the sum throughput active RIS-assisted uplink NOMA communication system. Reference \cite{Active_SWIPT} investigated active RIS-assisted simultaneous wireless information and power transfer (SWIPT) system, where the weighted sum-power harvested by energy users and the weighted sum-rate of the information users was maximized respectively by jointly optimizing transmit beamforming and active RIS beamforming. Reference \cite{Wideband} proposed wideband active RIS architecture. Reference \cite{Nearly_Passive} considered active RIS with global reflection constraints assisted wireless communication by jointly optimizing the transmit power of users, received beamforming of base station (BS), and RIS reflection coefficients maximizing global energy efficiency.
\subsubsection{Active vs Passive RIS-Assisted Communication: Performance Analysis}    
In reference \cite{Same_Power_Budget, Active_Passive, Reflect_or_Not}, the authors explored the potential limitations of active RIS-assisted communication. In \cite{Same_Power_Budget, Active_Passive}, the authors compared the throughput performance of passive RIS-assisted communication systems and active RIS-assisted communication systems, respectively. Their findings indicated that active RIS-assisted communication systems outperformed passive only when the number of reflecting elements was limited and sufficient energy was available. Furthermore, in \cite{Active_Passive}, the authors also discovered that when the amplification power of active RIS was limited and the RIS was deployed close to the transmitter, the benefits of amplified signals cannot compensate for amplification noise due to the low amplification factor of each reflecting element, making the performance of passive RIS-assisted communication systems superior to that of active RIS-assisted systems. In \cite{Reflect_or_Not}, it was found that for active RIS-assisted communication systems, when aiming to maximize the system's energy efficiency, some of the reflecting elements should be operated in idle mode, rather than all reflecting elements operating in active mode. 
\subsubsection{Research In Hybrid RIS}
Due to the limitations of active RIS, scholars have studied communication systems assisted by hybrid RIS \cite{New_Hybrid, Exploring_Hybrid}. Reference \cite{New_Hybrid} investigated hybrid RIS-assisted multi-user transmission systems in Rayleigh channel environments, where the ergodic capacity of the worst-case user was maximized by jointly optimizing the allocation of active/passive elements and beamforming of active/passive elements. Simulation results showed that hybrid RIS can effectively enhance the ergodic capacity of the worst-case user by integrating the advantages of both active and passive RIS. Reference \cite{Exploring_Hybrid} proposed a general hybrid RIS architecture, where each reflecting element can flexibly switch between active and passive modes. Simulation results showed the hybrid RIS can effectively reduce the trade-off between system energy consumption and latency compared to active RIS and passive RIS. 
\subsubsection{Research In Wireless Powered Passive RIS}
The research above assumed that RIS was powered by batteries or connected to the power grid. References \cite{Wireless-Powered,Optimized-Energy,TCCN,Robust-and-Secure,Energy-Harvesting,Computation-Offloading,Performance-Analysis,RIS-Aided-EH-NOMA,Multi-Functional} investigated communication systems assisted by wirelessly powered passive \cite{Wireless-Powered,Optimized-Energy,TCCN,Robust-and-Secure,Energy-Harvesting,Computation-Offloading}, active \cite{Performance-Analysis,RIS-Aided-EH-NOMA}, and hybrid RIS \cite{Multi-Functional}. 

Reference \cite{Wireless-Powered} investigated a multi-input single-output (MISO) system assisted by wireless-powered passive RIS (WPPR), where information
transmission was divided into two phases. In the first phase, the RIS harvested energy from  Access Point (AP) beamforming; in the second phase, RIS consumed energy to reflect the incident signal assisting information
transmission from the AP to the user. The system's sum throughput was maximized by optimizing the transmit beamforming, time allocation, and RIS phase shifts subject to the maximum transmit power of AP and available energy of  RIS. Simulation results showed that a WPPR-assisted communication system can significantly improve the sum throughput compared to schemes without RIS. Reference \cite{Optimized-Energy} investigated WPPR-assisted WPCN, where RIS utilized time splitting (TS) and power splitting (PS) methods to harvest energy from RF signals transmitted by the hybrid access point (HAP) realizing self-sustainability. The sum rate of the system was maximized by optimizing the transmission power of user terminals, time allocation, RIS phase shifts, and amplitude reflection coefficients subject to the energy constraints of both RIS and users. Simulation results showed that WPPR can significantly increase the system sum rate compared to the random phase shift RIS scheme. In \cite{TCCN}, the AP’s transmit power was minimized in the WPPR-assisted MISO system, where RIS harvested energy by PS method, by optimizing the transmission power of user terminals, RIS phase shifts, and reflection coefficients, constrained by the available energy of both RIS and receiver’s SNR requirements. The results showed that the WPPR can significantly reduce the energy consumption of the AP compared to schemes without RIS. Reference \cite{Robust-and-Secure} investigated WPPR-assisted multiuser MISO downlink communications, where RIS harvested RF energy by switching a portion of elements in power harvest mode. With uncertain channel information, the system's sum rate was maximized by optimizing transmission beamforming, RIS phase shifts, and the operating modes of the RIS element, constrained by the maximum transmit power budget of AP, available energy of  RIS, and maximum tolerable information leakage. The simulation results unveiled that compared to non-RIS-assisted communication, WPPR-assisted communication significantly increased the sum rate of the system. Reference \cite{Energy-Harvesting} investigated UAV-assisted wireless information and power transfer (SWIPT) system, where WPPR was installed on UAV and each reflecting element can operate in either energy harvesting or passive mode. Deep reinforcement learning was proposed to maximize the system's energy efficiency by optimizing time allocation, the operating modes of reflecting elements, RIS phase shifts, and BS transmission beamforming given BS transmission power constraints and QoS constraints. In \cite{Computation-Offloading}, a WPPR-assisted edge computing system consisting of a HAP and multiple user terminals was studied. An Optimization-driven hierarchical reinforcement learning optimization algorithm was proposed to minimize the system's total energy consumption by optimizing RIS phase shifts, time allocation, task offloading ratios, user transmission power, and HAP beamforming. 
\subsubsection{Research in wireless powered active RIS}
Reference \cite{Performance-Analysis,RIS-Aided-EH-NOMA} focused on the wireless powered active RIS (WPAR) assisted communication system. In reference \cite{Performance-Analysis}, authors studied the downlink transmission in a WPAR-assisted communication system consisting of a BS and multiple users, where each reflecting element can operate in either active mode or energy harvesting mode. With uncertain channel
information, the sum rate was maximized by optimizing the operating mode of each RIS reflecting element, BS transmit beamforming, RIS phase shift, and the amplification factor of active elements. The simulation results revealed that when the transmit power of BS was sufficiently large, the thoughtout performance gap between WPAR and grid-powered passive RIS is negligible. In \cite{RIS-Aided-EH-NOMA}, the WPAR-assisted downlink NOMA system was investigated, where the long short-term memory-deep deterministic policy gradient was proposed to maximize the communication success ratio by jointly optimizing active RIS amplification factor and phase shift. 
\subsubsection{Research In Wireless Powered Hybrid RIS}
Reference \cite{Multi-Functional} focused on hybrid RIS-assisted downlink NOMA wireless communication system, where each reflecting element has the functionalities of signal reflection, transmission, amplification, and energy harvest. The sum rate was maximized by optimizing each reflecting element's operating mode and the beamforming of BS and RIS. Simulation results showed that the hybrid RIS-assisted communication scheme can significantly improve the system's throughput compared to the WPPR-assisted communication scheme.
\subsection{Motivations and Contributions}

Owing to the advantages of hybrid RIS-assisted communication, and the flexibility of wireless RF energy harvesting technology in addressing energy limitations of wireless communication networks, this paper proposes research on wireless-powered hybrid RIS-assisted communication systems. Each reflecting element can operate in active, passive, or idle mode in these systems. Since reference \cite{Multi-Functional} already investigated wireless-powered hybrid RIS-assisted communication systems, the most relevant work to this research is that of reference \cite{Multi-Functional}. However, the signal model proposed in reference \cite{Multi-Functional} is relatively simple and rough, unable to accurately describe the received signal at the receiver when some reflecting elements are operating in passive mode.

Furthermore, reference \cite{Multi-Functional} only considered the scenario of downlink communication, where hybrid RIS harvested energy from BS beamforming by operating a portion
of the element in energy harvesting mode. However, for uplink communications, where source nodes are user terminals or sensor nodes with limited transmission power, the energy harvested by the wireless-powered RIS from RF signals transmitted by the source nodes is negligible. This limitation renders the findings of reference \cite{Multi-Functional}
inapplicable to the uplink communication assisted by wireless-powered hybrid RIS. 

Two shortcomings mentioned above  motivate the work of this paper. In this paper, we investigate a new scheme for a wireless-powered hybrid RIS-assisted uplink communication system, where the RIS assists multiple users in uploading information to BS by TDMA
protocol. The contributions of our work can be summarized as follows:

\begin{itemize}
\item We propose a novel system model for wireless-powered hybrid RIS-assisted uplink communication. This system integrates EHS within the RIS to harvest energy from the RF signals transmitted by ES. Considering that active reflection can overcome the ’multiplicative fading’ effect and passive reflection offers the benefit of low power consumption, and further inspired by reference \cite{Reflect_or_Not} which indicates that operating portion of the reflecting element in idle mode can improve energy efficiency, thus this paper investigates hybrid RIS, where each reflecting element can operate in either passive, active, or idle mode. 
\end{itemize}

\begin{itemize}
\item We derive a more accurate signal model that can express the received signal at the receiver when the reflected element operates in passive mode and idle mode.
\end{itemize}

\begin{itemize}
\item We introduce the TS energy harvest method, which divides the transmission block into two phases. In the first phase, Energy harvest surface (EHS) harvests energy and stores it in rechargeable batteries. In the second phase, EHS continues to harvest energy
while RIS utilizes the energy from the rechargeable batteries and the energy harvested by the EHS to reflect incident signals. In this way, the RIS can assist communication by an appropriate number of reflective elements operating in active and passive modes to achieve optimal system performance.
\end{itemize}

\begin{itemize}
\item A min-max energy consumption problem with the joint design of the operating modes of each reflecting element, the amplification factor of active elements, the transmit power, and transmission time allocation are investigated by considering QoS constrain as well as available energy of RIS constrain. Owing to the operational modes of each reflecting element being described by discrete optimization variables and being highly coupled with the amplification factors, time allocation, and the transmit power of each user's terminal, the optimization problem formulated is a challenging mixed-integer programming (MIP) problem. Consequently, this paper proposes a hierarchical optimization approach based on deep reinforcement learning to solve this problem. In this approach, the operating modes of each reflecting element and the amplification factors are solved in the outer subproblem by deep reinforcement learning methods, while time allocation and the transmit power of each user are solved in the inner subproblem by convex optimization methods.
\end{itemize}

\begin{itemize}
\item Simulation results show that the proposed optimization scheme can effectively reduce the energy consumption of user terminals compared with schemes without RIS. Furthermore, the simulation results also reveal that the hybrid RIS tends to extend the RIS operating time and increase the amplification factor of active elements by increasing the number of idle elements to minimize the maximum energy consumption of the users.
\end{itemize}

\section{System model}
As shown in Figure 1, we consider the uplink transmission in a wireless-powered hybrid RIS-assisted communication system consisting of an ES with a single antenna, a RIS, a single-antenna BS, and J single-antenna user terminals. There is a direct link between the user terminals and the BS. Still, the direct link is weak due to obstructions, necessitating an RIS installed on the walls of buildings to assist communication between them. The RIS is assumed to be equipped with $N$ reflecting elements, each of which can operate in passive mode, active mode, or idle mode \cite{Reflect_or_Not}. In idle mode, the reflecting element is inactive and does not consume energy. Furthermore, to ensure sufficient energy is available for the reflecting elements to operate in active reflection mode, the system includes EHS embedded with $M$ energy harvesting element \cite{Polarization-Independent}. This surface harvests energy from the RF transmitted by the ES, storing the harvested energy in a rechargeable battery to power the RIS circuits.

Let $\beta_n$ be the indicator that determines whether the $n$-th reflecting element operates in signal reflection mode. In other words, when $\beta_n=1$ the $n$-th reflecting element operates in either passive mode or active mode; otherwise, it operates in idle mode. Furthermore, When $\beta_n$ = 1, let $\alpha_n$ be the signal reflection mode indicator for the $n$-th reflecting element. Specifically, when $\beta_n=1$ and $\alpha_{n}=1$, the $n$-th reflecting element operates in active mode; conversely, when $\beta_n=1$ and $\alpha_{n}=0$, it operates in passive mode. Motivated by \cite{Exploring_Hybrid,Reflect_or_Not}, the reflection coefficient matrix of the RIS can be defined as $\boldsymbol{\gamma}=\operatorname{diag}\left(\beta_1 \rho_1^{\alpha_1} e^{j \theta_1}, \cdots, \beta_n \rho_n^{\alpha_n} e^{j \theta_n}, \cdots, \beta_N \rho_N^{\alpha_N } e^{j \theta_N}\right)$, where $\theta_n \in[0,2 \pi]$ denotes the phase shift of the $n$-th reflecting element, and $\rho_n \geq 1$ denotes the amplification factor when the $n$-th reflecting element operates in active mode. Note that when $\beta_n=0$, the $n$-th element on the diagonal of $\boldsymbol{\gamma}$ equal to 0, indicating that the $n$-th reflecting element operates in idle mode.
 \begin{figure*}[ht]
\centering
\includegraphics[width=6.5in]{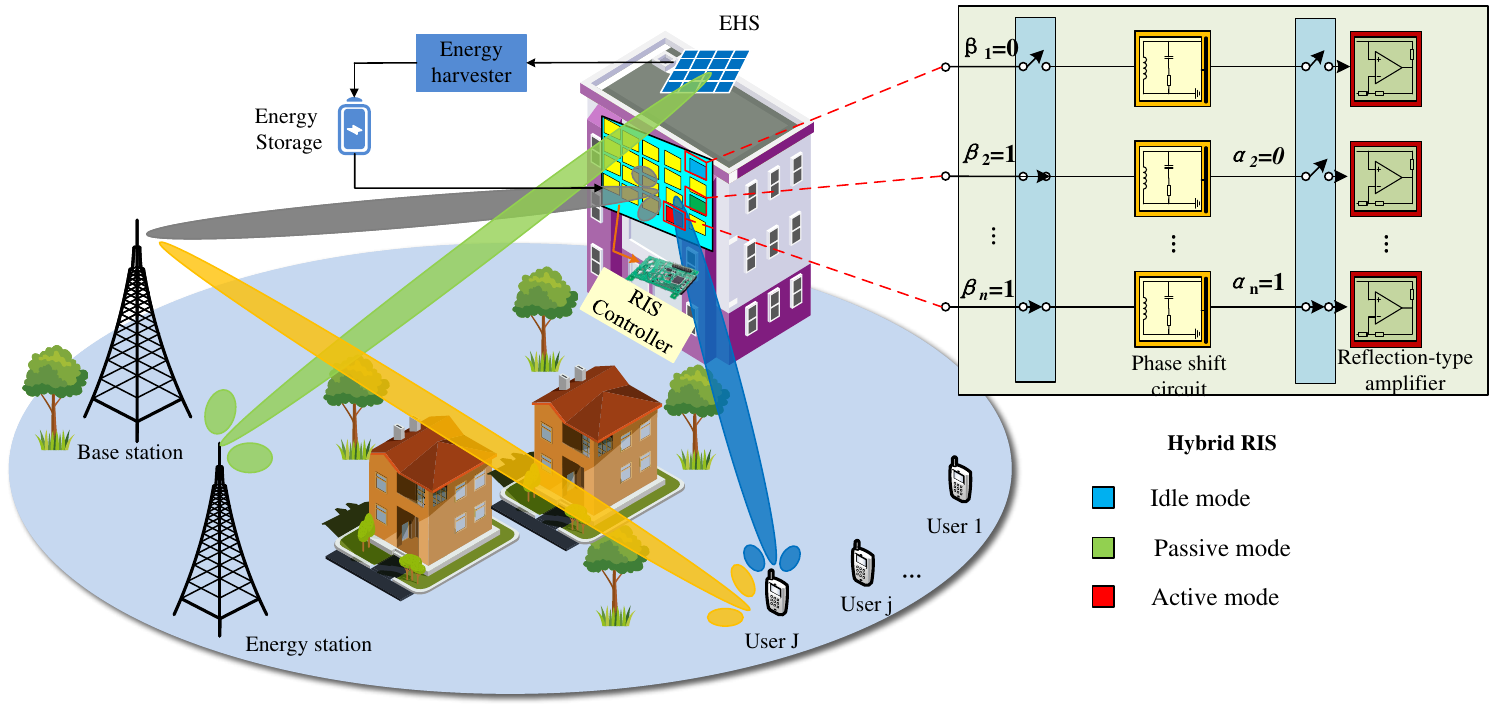}
\centering
\caption{System model}
\end{figure*}
\printcredits

\begin{table}[pos=H] %
\captionsetup{justification=centering}
\caption{\centering NOTATION DEFINITIONS}
\begin{tabular}{m{2cm}<{\centering}|m{5cm}<{\centering}} % 
\hline
\hline
 {\bf \small Symbol}&\qquad {\bf\small Description} \\ 
\hline 
$\mathcal{J}, J, j$ & The set, quantity, and indices of user terminals \\
\hline
$t_{i, j}$ & The duration of time for which user terminal $j$ sends information during the $i$ th phase \\
\hline
$\mathbb{N}, N, n$ & The set, quantity, and indices of RIS elements \\
\hline
$\alpha_n$ & Indicator of the operating mode for the $n$-th element, identified as either active mode or passive mode.\\
\hline
$\beta_n$ & Indicator of the operating mode for the $n$-th element, identified as either signal reflection mode or idle mode. \\
\hline
$\rho_n$ & The amplification factor of the $n$-th reflective element when operating in active mode.  \\
\hline
$P_{\mathrm{C}}$ & the power consumption of phase-shift circuit\\
\hline
$P_{\mathrm{DC}}$ & the power consumption of amplifier circuit\\
\hline
$P_b$ & the power consumption of the
RF-to-DC power conversion circuit \\
\hline
$Q_{\text {min }}$ & the minimum required transmission
rate of each user \\
\hline
\hline
\end{tabular}
\end{table}

For ease of reading, Table 1 summarizes the description of the symbols used in the remainder of this paper.
\subsection{Protocol for Wireless-Powered RIS-Assisted Communication}
Fig. 2 shows the transmission protocol for the considering system. According to Figure 2, the transmission block is divided into two phases. In the first phase, the ES transmits RF signals, and the RIS harvests energy from it through EHS, storing the harvested energy in a rechargeable battery. Meanwhile, $J$ users send information to the BS via direct links using TDMA protocol. Denote the duration of upload information in the first phase for $j$-th user is $t_{1,j}$. In the second phase, while the ES continues to transmit RF signals, and $J$ users continue to send information using TDMA protocol, the RIS consumes the energy harvested by the EHS from two phases to assist users in uploading information to BS. In other words, each user can send information to the BS simultaneously through both the direct link and the reflecting link in the second phase. Denote the duration of upload information in the second phase for $j$-th user is $t_{2,j}$. Considering that the RF signals from the ES are known to the BS, it is assumed here that the BS can eliminate the influence of the RF signals transmitted by the ES on decoding user information in both phases by using self-interference cancellation \cite{Hybrid-SWIPT}.

 \begin{figure}[ht]
\centering
\includegraphics[width=3.3in]{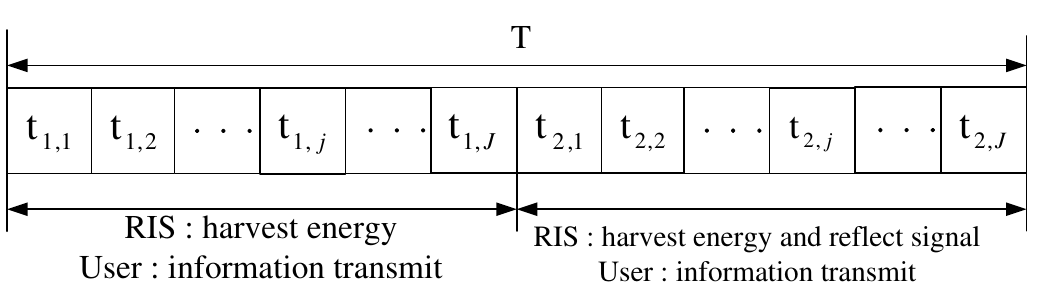}
\centering
\caption{The structure of time frame}
\end{figure}

According to the frame structure in Figure 2, the time allocation for each user in the two phases needs to satisfy the following inequality:
\begin{equation}\label{t1}
\sum_{i=1}^2 \sum_{j=1}^J t_{i, j} \leq T
\end{equation}
and 
\begin{equation}\label{t2}
t_{i, j} \geq 0, \forall i \in\{1,2\}, j \in \mathcal{J}
\end{equation}
where $\mathcal{J} \triangleq\{1, \cdots, J\}$ and $T$ denotes the during of time frame.

\subsection{Channel Model}

For the considered system, let the direct link channel coefficient from user terminal $j$ to BS, from user terminal $j$ to RIS, from RIS to BS, from ES to EHS are denoted as $h_j^{\mathrm{UB}} \in \mathbb{C}$, $\boldsymbol{h}_j^{\mathrm{UR}} \in \mathbb{C}^{N \times 1}$, $\boldsymbol{h}^{\mathrm{RB}} \in \mathbb{C}^{N \times 1}$, $\boldsymbol{h}^{\mathrm{PE}} \in \mathbb{C}^{M \times 1}$, respectively. It is assumed that these channels follow block-fading, meaning the channel coefficients remain constant within one-time slot but may independently change across different time slots. Due to the presence of obstacles between the user terminals and the BS, $h_j^{U B}$ is modeled by Rayleigh fading, which can be expressed as
\begin{equation}
h_j^\mathrm{U B}=\sqrt{\varrho_{0}\left(D_j^\mathrm{U B}\right)^{-\alpha_j^\mathrm{U B}}} \tilde{h}_j^\mathrm{U B}, \quad \forall j \in \mathcal{J}
\end{equation}
where $\varrho_{0}$ denotes path loss at the reference distance of 1$m$, $D_j^\mathrm{U B}$ denotes the distance between user terminal $j$ and BS, $\alpha_j^\mathrm{U B}$ denotes path loss exponents of the channel from user terminal $j$ to the BS and $\tilde{h}_j^\mathrm{U B}$  denotes a random variable which follows a complex Gaussian distribution, i.e., $\tilde{h}_j^\mathrm{U B} \sim \mathcal{C} \mathcal{N}\left(0, \sigma^2 \right)$. It is assumed that there are no obstructions between the user terminal and the RIS, between the RIS and the BS, and between the ES and the EHS. Therefore,  the channel coefficient vectors $\boldsymbol{h}_j^\mathrm{U B}$, $\boldsymbol{h}^\mathrm{RB}$, and $\boldsymbol{h}^\mathrm{ES}$ are modeled by Ricean fading. $\boldsymbol{h}_j^\mathrm{UR}$ is expressed as:
\begin{align}
\boldsymbol{h}_j^\mathrm{UR}= & \sqrt{\varrho_0\left(D_{j}^{\mathrm{UR}}\right)^{-\alpha_{j}^\mathrm{UR}}}\left(\sqrt{\frac{K_{j}^\mathrm{UR}}{K_{j}^\mathrm{UR}+1}} \boldsymbol{h}_{j,\mathrm{L O S}}^\mathrm{UR}+ \right. \nonumber \\
& \left.\sqrt{\frac{1}{K_{j}^\mathrm{UR}+1}} \boldsymbol{h}_{j,\mathrm{N L O S}}^\mathrm{UR}\right), \quad \forall j \in \mathcal{J}
\end{align}
where $K_{j}^{\mathrm{UR}}$ is Ricean fading, $D_{j}^{\mathrm{UR}}$ and $\alpha_{j}^{\mathrm{UR}}$ denote the distance from user terminals $j$ to RIS and path loss exponents of the channel from user terminals $j$ to RIS. For a RIS configured as a uniform linear array (ULA), the line-of-sight (LOS) component $\boldsymbol{h}_{j, \mathrm{L O S}}^\mathrm{UR}$ can be described as the array response of N elements ULA. Thus, $\boldsymbol{h}_{j, \mathrm{L O S}}^\mathrm{UR}$ is given by:
\begin{equation}
\left[1, e^{-j \frac{2 \pi}{\lambda} d_R \phi_j^\mathrm{UR}}, \ldots, e^{-j \frac{2 \pi}{\lambda}(N-1) d_R \phi_j^\mathrm{UR}}\right]^T
\end{equation}
$\phi_j^\mathrm{UR}$ denote the angle of
departure (AoD) of the signal, $d_{R}$ is the distance between adjacent reflecting elements, and $\lambda$ is the wavelength of the signal. The non-line-of-sight (NLOS) component $\boldsymbol{h}_{j,\mathrm{NLOS}}^{\mathrm{UR}}$ is a random variable that follows a circularly symmetric complex Gaussian distribution, i.e., $\boldsymbol{h}_{j,\mathrm{NLOS}}^{\mathrm{UR}} \sim \mathcal{C} \mathcal{N}\left(\mathbf{0}, \sigma^2 \mathbf{I}_N\right)$. The channel coefficient vectors $\boldsymbol{h}^{\mathrm{RB}}$ and $\boldsymbol{h}^{\mathrm{ES}}$ can be uniformly expressed as:
\begin{align}
\mathbf{h}^\mathrm{S}= & \sqrt{\varrho_0\left(D^\mathrm{S}\right)^{-\alpha^\mathrm{S}}}\left(\sqrt{\frac{K_\mathrm{S}}{K_\mathrm{S}+1}} \boldsymbol{h}_\mathrm{L o S}^\mathrm{S}+ \right. \nonumber \\
& \left. \sqrt{\frac{1}{K_\mathrm{S}+1}} \boldsymbol{h}_{\mathrm{NLoS }}^\mathrm{S}\right), \mathrm{S} \in\{\mathrm{UR}, \mathrm{ES}\}    
\end{align}
where $D^\mathrm{S}$ and $\alpha^\mathrm{S}$ denote the distance and path loss exponents of the channel, respectively. The Rice factor for these channels is denoted as $K_\mathrm{S}$. The terms $h_{\mathrm{LOS}}^\mathrm{S}$ and $h_{\mathrm{NLOS}}^\mathrm{S}$ refer to LoS component and NLoS component, respectively. $h_{\mathrm{NLOS}}^\mathrm{S}$ denotes a random variable following a circularly symmetric complex Gaussian distribution, while $\boldsymbol{h}_{\mathrm{L O S}}^\mathrm{S}$ can be expressed as:
\begin{equation}
\left[1, e^{-j \frac{2 \pi}{\lambda} d_\mathrm{s} \phi^\mathrm{s}}, \ldots, e^{-j \frac{2 \pi}{\lambda}(N-1) d_\mathrm{s} \phi^\mathrm{s}}\right]
\end{equation}
where $\phi^\mathrm{s}$ denotes the angle of departure (AoD) or AOA of the signal. For the RB channel, $d_{\mathrm{s}}=d_{\mathrm{R}}$; for the PE channel, $d_{\mathrm{s}}= d_{\mathrm{E}}$, where $d_{\mathrm{E}}$ denotes the distance between adjacent energy harvesting elements on EHS.

\subsection{Signal Model}
In the first phase, the BS receives signals sent by users only through the direct link, but in the second phase, the BS can receive signals sent by users through both the direct link and the reflecting link. Let the signal sent by the $j$-th user in the $i$-th phase be $s_{i, j}$, where $s_{i, j}$ is an independent complex Gaussian random variable with zero mean and unit variance. The transmit power of the $j$-th user in the $i$-th phase is denoted as $p_{i,j}$. Therefore, when user $j$ sends information in the first phase, the signal received by BS can be expressed as:
\begin{equation}\label{direct}
y_{1, j}=h_j^{\mathrm{U B}} \sqrt{p_{1, j}} s_{1, j}+z_\mathrm{B}
\end{equation}
where $z_\mathrm{B} \sim \mathcal{C N}\left(0, \sigma_\mathrm{B}^2\right)$ denotes the additive white Gaussian noise (AWGN) at the BS.

In the second phase, each reflecting elements can operate in passive mode, active mode, or idle mode. To avoid additional power consumption caused by frequent mode switching, this paper assumes that the operating mode of each reflective element remains unchanged during different time slots within a time frame, but the phrase can change flexibly. Therefore, the reflection coefficient matrix during $t_{2,j}$, can be expressed as

\begin{align}
\boldsymbol{\gamma_j}=\operatorname{diag}\left(\beta_1 \rho_1^{\alpha_1} e^{j \theta_{1, j}}, \cdots, \beta_N \rho_N^{\alpha_N} e^{j \theta_{N, j}}\right)    
\end{align}
if the reflecting elements operate in active mode, the thermal noise will be introduced and amplified. Conversely, when operating in passive or idle mode, thermal noise is not generated. Motivated by \cite{Exploring_Hybrid,Reflect_or_Not}, the noise amplification coefficient matrix can be defined as:

\begin{equation}
\boldsymbol{\Phi_j}=\operatorname{diag}\left(\alpha_1 \beta_1 \rho_1 e^{j \theta_{1, j}}, \cdots,
\alpha_N \beta_N \rho_N e^{j \theta_{N, j}}\right)  
\end{equation}
The optimal phase of the $n$-th reflecting element for user $j$ can be expressed as \cite{Active_first, Optimized-Energy}:
\begin{equation}
\theta_{n, j}^*=\arg \left(h_j^{\mathrm{U B}}\right)-\arg \left(\left(\boldsymbol{h}_n^{\mathrm{R B}}\right)^H\right)-\arg \left(\boldsymbol{h}_{j,n}^{\mathrm{U R}}\right)
\end{equation}
where $\boldsymbol{h}_n^{\mathrm{R B}}$ is the $n$-th element of $\boldsymbol{h}^{\mathrm{R B}}$, $\boldsymbol{h}_{j, n}^{\mathrm{UR}}$ is the $n$-th element of $\boldsymbol{h}_{j}^{\mathrm{UR}}$ and $\arg (x)$ is phrase of $x$. Therefore, when user $j$ sends information in the second phase, the signal received by BS can be expressed as:
\begin{align}\label{reflect}
y_{2, j}&=\underbrace{{h}_j^{\mathrm{U B}} \sqrt{p_{2, j}} s_{2, j}}_{\text {direct link }}\nonumber\\
&+\underbrace{\left(\boldsymbol{h}^{\mathrm{R B}}\right)^H \boldsymbol{\gamma_j} \boldsymbol{h}_j^{\mathrm{U R}} \sqrt{p_{2, j}} s_{2,j}+\left(\boldsymbol{h}^{\mathrm{R B}}\right)^H \boldsymbol{\Phi_j} \mathbf{z}_\mathrm{R}}_{R I S \text {-aided link }}\nonumber\\
&+z_\mathrm{B}
\end{align}
where $\mathbf{z}_\mathrm{R} \in \mathbb{C}^{N \times 1}$ denotes the thermal noise introduced by active element, which is follows a circularly symmetric complex Gaussian distribution, i.e., $\mathbf{z}_R \sim \mathcal{C} \mathcal{N}\left(\mathbf{0}, \sigma_{F}^2 \mathbf{I}_N\right)$. In this paper, it is assumed that each user must upload at least $Q_{min}$ bits of information to the BS within a single time frame. Therefore, based on equations \eqref{direct} and \eqref{reflect}, there exists the following Quality of Service (QoS) constraint:
\begin{align}\label{QoS}
&t_{2, j} \log _2\left(1+\frac{p_{2, j}\left|h_j^{\mathrm{UB}}+\left(\boldsymbol{h}^{\mathrm{R B}}\right)^H \boldsymbol{\gamma_j} \boldsymbol{h}_j^{\mathrm{U R}}\right|^2}{\sigma_{\mathrm{B}}^2+\sigma_{\mathrm{F}}^2\left\|\left(\boldsymbol{h}^{\mathrm{R B}}\right)^H \boldsymbol{\Phi_j}\right\|^2}\right) \nonumber\\
&+t_{1, j}\log _2\left(1+\frac{p_{1, j}\left|h_j^{\mathrm{UB}}\right|^2}{\sigma_{\mathrm{B}}^2}\right) \geq Q_{\min }, \forall j \in \mathcal{J}
\end{align}
\subsection{Energy model}
According to the frame structure shown in Figure 2, the energy consumed by passive elements and active elements during the second phase can be respectively expressed as:
\begin{equation}
E^{p a s}=\sum_{n=1}^N\left(1-\alpha_n\right) \beta_n P_{\mathrm{C}}\left(\sum_{j=1}^J t_{2, j}\right)
\end{equation}
\begin{align}
&E^{a c t}=\sum_{n=1}^N \alpha_n \beta_n\left(P_{\mathrm{C}}+P_{\mathrm{DC}}\right) \left(\sum_{j=1}^J t_{2, j}\right) \nonumber\\
&+\xi\left(\sum_{j=1}^J\left\| \boldsymbol{\Phi_j} \boldsymbol{h}_j^{\mathrm{U R}}\right\|^2 p_{2, j} t_{2, j}+\sum_{j=1}^J \sigma_{\mathrm{F}}^2\left\| \boldsymbol{\Phi_j}\right\|^2 t_{2, j}\right) 
\end{align}
where $P_{\mathrm{C}}$ denotes the power consumption of
phase-shift circuit, $P_{D C}$ denotes the power consumption of amplifier circuit and $\xi$ denotes the inverse of amplifier efficiency.
The energy consumption for EHS in a time frame can be expressed as:
\begin{equation}
E^{\mathrm{R F-D C}}=M P_\mathrm{b} T
\end{equation}
where $P_b$ denotes the power consumption of the RF-to-DC power
conversion circuit.
The energy consumed by the RIS and EHS within a time frame must not exceed the energy harvested by the EHS within a time frame, therefore we have following energy constrain:
\begin{equation}\label{energy_constrain}
E^{\mathrm{p a s}}+E^{\mathrm{a c t}}+E^{\mathrm{R F-D C}} \leq \eta p_s\left\|\boldsymbol{h}^{\mathrm{ES}}\right\|^2 T
\end{equation}
where $\eta$ is the EH efficiency and $p_{s}$ is  transmit power of ES. 
\subsection{Problem Formulating}
In this paper, our goal is to maximize the minimum energy consumption of by jointly optimizing the operating modes of each reflecting element, the amplification factor of active elements, the transmit power and transmission time allocation, subject to quality-of-service (QoS) of each user and the available energy constraint of RIS. Therefore, the
corresponding optimization problem can be formulated as
follows:
\begin{subequations}\label{zong}
\begin{align}
\label{zong1}
(\mathrm{P} 1): &\min _{\left\{p_{i, j}, t_{i, j}, \alpha_n, \beta_n, \rho_n\right\}}\left\{\max _{j \in \mathcal{J}} \sum_{i=1}^2 p_{i, j} t_{i, j}\right\}\\
\label{zong2}
&\alpha_n \in\{0,1\}, \beta_n \in\{0,1\}, \forall n \in \mathbb{N}\\
\label{zong3}
&\eqref{t1},\eqref{t2},\eqref{QoS},\eqref{energy_constrain}\\
\label{zong4}
&0 \leq p_{i, j} \leq p_{\max }, \forall i \in\{1,2\}, j \in \mathcal{J}\\
\label{zong5}
&0<\rho_{\mathrm{n}} \leq \rho_n^{\max }
\end{align}
\end{subequations}
where $p_{\max }$ denotes maximum transmit power of user and $\rho_n^{\max }$ denotes maximum amplitude of active elements. \eqref{zong2} is binary constraint about modes operating 
of each reflecting element. \eqref{zong4} is constrains on user maximum transmit power. \eqref{zong5} is the constrains on amplification factors. Problem (P1) is a challenging mixed-integer programming(MIP) problem. Moreover, in constraints \eqref{QoS} and \eqref{energy_constrain}, the binary optimization variables $\alpha_n$ and $\beta_n$ are coupled with the continuous variables $\rho_{\mathrm{n}}$, $t_{i, j}$, and $p_{i, j}$, further increasing the difficulty of solving the problem. However, it can be observed that when the operation mode of reflecting elements and the amplification factor are given (i.e., $\alpha_{n}$, $\beta_{n}$, and $\rho_{\mathrm{n}}$ are given), Problem P1 can be simplified as:

\begin{subequations}\label{zong-2}
\begin{align}
&\mathop (\mathrm{P}2):F^*(\boldsymbol{\alpha}, \boldsymbol{\beta}, \boldsymbol{\rho})=\min _{\left\{E_{i, j}, t_{i, j}\right\}}\left\{\max _{j \in \mathcal{J}} \sum_{i=1}^2 E_{i, j}\right\}\\
&\eqref{t1},\eqref{t2}\\
\label{Q_min}
&t_{2, j} \log _2\left(1+\frac{E_{2, j}\left|h_j^\mathrm{U B}+\left(\boldsymbol{h}^{\mathrm{R B}}\right)^H \boldsymbol{\gamma_j} \boldsymbol{h}_j^{\mathrm{U R}}\right|^2}{t_{2, j}\left(\sigma_{\mathrm{B}}^2+\sigma_{\mathrm{F}}^2\left\|\left(\boldsymbol{h}^{\mathrm{R B}}\right)^H \boldsymbol{\Phi_j}\right\|^2\right)}\right)  \nonumber\\
&+ t_{1, j} \log _2\left(1+\frac{E_{1, j}\left|h_j^{\mathrm{U B}}\right|^2}{t_{1, j} \sigma_{\mathrm{B}}^2}\right) \geq Q_{\min }, \forall j \in \mathrm{J}\\
&0 \leq E_{i, j} \leq p_{\max } t_{i, j}, \forall i \in\{1,2\}, j \in \mathrm{J}\\
&\xi\left(\sum_{j=1}^J\left\|\boldsymbol{\Phi_j} \boldsymbol{h}_j^{\mathrm{UR}}\right\|^2 E_{2, j}+\sum_{j=1}^J \sigma_{\mathrm{F}}^2\left\|\boldsymbol{\Phi_j}\right\|^2 t_{2, j}\right) \nonumber\\
&+\sum_{n=1}^N  \alpha_n \beta_n\left(P_\mathrm{C}+P_{\mathrm{D C}}\right)\left(\sum_{j=1}^J t_{2, j}\right)+E^{\mathrm{R F-D C}} \nonumber\\
&+E^{\mathrm{p a s}} \leq \eta p_s\left\|\boldsymbol{h}^{\mathrm{ES}}\right\|^2 T
\end{align}
\end{subequations}
where $E_{i, j}=p_{i, j}t_{i,j}$. It can be proven that Problem (P2) is a convex problem and can be effectively solved by the interior point method. Based on this, we propose a hierarchical optimization method based on deep reinforcement learning (DRL) to solve the problem (P2) by decomposing the problem (P1) into outer sub-problem and the inner sub-problem (P2). For the outer sub-problem, we reformulate it as a Markov Decision Process (MDP) and DRL is used to learn to output the optimal operating modes of each reflecting elements and the amplification factor (i.e., $\left\{\alpha_n, \beta_n, \rho_n \mid n \in \mathbb{N}\right\}$) based on input channel station; for the inner sub-problem (P2), the interior point method is used to solve Problem P2 to obtain optimal transmit power for each user and time allocation(i.e., $\left\{p_{i, j}, t_{i, j} \mid i \in\{1,2\}, j \in \mathcal{J}\right\}$) given $\left\{\alpha_n, \beta_n, \rho_n \mid n \in \mathbb{N}\right\}$. The specific details will be described in the next section.

\section{Proposed solution: PPO-Convex}
Since the outer sub-problem involves both binary and continuous variables, the corresponding action space of the MDP problem is a discrete-continuous hybrid action space. For such problems, a common approach is to discretize the continuous actions and then solve them by the Deep Q-network (DQN) \cite{Hierarchical-Reinforcement}. However, DQN is not suitable for solving the outer sub-problems. The reason is that for an RIS with $N$ reflecting elements, the action space size for the operating mode (i.e.,$\left\{\alpha_n, \beta_n \mid n \in \mathbb{N}\right\}$ ) is $4^{N}$. As the number of reflecting elements increases, the size of the action space grows exponentially, hindering the convergence of training for DQN. Conversely, the PPO method can effectively address MDP problems with large-scale action spaces \cite{Gym-µRTS}. Therefore, this section will propose an efficient algorithm based on the PPO method to solve the outer sub-problems corresponding to Problem P1.

In this section, we will introduce preliminary knowledge of the PPO algorithm, including the theoretical basis of policy gradient (PG) algorithms and the PPO algorithm. Then, we will formulate the corresponding MDP problem based on the outer sub-problem. Finally, we will elaborate on how to solve the problem P1 using a hierarchical optimization approach based on DRL.

\subsection{Preliminary Knowledge}
\subsubsection{PG algorithms}

The essence of policy gradient algorithms is to enable agents to make decisions $a_{t}$ based on different environments $s_{t}$ in order to maximize the cumulative reward $R_t$. At the initial time step $t_{1}$
 , the agent determines an action $a_{1}$ according to the current state $s_{1}$ by the policy function $\pi\left(a_t \mid s_t\right)$. This action leads the environment to transition from state $s_{1}$ to $s_{2}$, granting the agent a reward $r_{1}$. Subsequently, the agent continues this process, selecting a new action $a_{2}$, transitioning to a new state $s_{3}$
, and receiving the corresponding reward $r_{2}$, and so on in a cycle, gradually forming an experience trajectory $\tau=\left\{s_t, a_t, s_t, a_t, \ldots \ldots, s_p, a_p\right\}$. Considering the strong nonlinear fitting capability demonstrated by Deep Neural Networks (DNNs), which allows them to approximate almost any complex function, using policy networks to approximate the policy function has become an effective approach in reinforcement learning. The state value function of the trajectory $\tau$
can be defined as:
\begin{align}
V_\pi\left(s_t\right)=\mathrm{E}_{A_t \sim \pi\left(\mid s_t ; \boldsymbol{\mu}\right)}\left[Q_\pi\left(s_t, A_t\right)\right]
\end{align}
It denotes the expected reward that an agent can obtain in state $s_{t}$ if it chooses an action following the policy network $\pi\left(a_t \mid s_t\right)$ with parameters $\boldsymbol{\mu}$, where $Q_\pi\left(s_t, a_t\right)$ is the action value function. This function indicates the expected reward that the agent can obtain when in state $s_{t}$, it takes action $a_{t}$ and subsequently selects the actions sampled by the policy network $\pi\left(a_t \mid s_t ; \boldsymbol{\mu}\right)$. $Q_\pi\left(s_t, a_t\right)$ can be expressed as:
\begin{equation}
Q_\pi\left(s_t, a_t\right)=\mathrm{E}_{s_{t+1}, s_{t+1}, \cdots, s_p, a_p}\left[U_t \mid S_t=s_t, A_t=a_t\right]
\end{equation}
where denotes $U_t$ discount reward, it can expressed as:
\begin{align}
U_t=\sum_{k=t}^p \gamma^{k-t} \cdot R_k
\end{align}
where $\gamma$ denote discount factor. The policy gradient algorithm is proposed to maximize the expected state-value function under the policy network 
$\pi\left(a_t \mid s_t ; \boldsymbol{\mu}\right)$, which is expressed as
\begin{align}
\max J(\boldsymbol{\mu})=\mathbb{E}_S\left[V_\pi\left(s_t\right)\right]
\end{align}
A common method for maximizing $J(\boldsymbol{\mu})$
is to perform gradient ascent. For a trajectory 
$\tau$, the gradient of $J(\boldsymbol{\mu})$ can be expressed as:
\begin{align}
\nabla J(\boldsymbol{\mu})=\sum_{k=t}^p Q_\pi\left(s_k, a_k\right) \cdot \nabla_{\boldsymbol{\mu}} \ln \pi\left(a_k \mid s_k ; \boldsymbol{\mu}\right)
\end{align}
where $Q_\pi\left(s_k, a_k\right)$ is estimated by $\sum_{t^{\prime}=k}^p \gamma^{\left(t^{\prime}-k\right)} r_{t^{\prime}}$, the update formula for the policy network can be expressed as:
\begin{align}
\boldsymbol{\mu} \leftarrow \boldsymbol{\mu}--\nabla J(\boldsymbol{\mu})
\end{align}

To reduce the variance of $\nabla J(\boldsymbol{\mu})$ during training and make the training converge faster, the advantage function $A_\pi\left(s_k, a_k\right)$  is used to replace $\sum_{t^{\prime}=k}^p \gamma^{t^{\prime}-k} r_{t^{\prime}}$, where $A_\pi\left(s_k, a_k\right)=\sum_{t^{\prime}=k}^p \gamma^{t^{\prime}-k} r_{t^{\prime}}-V_\pi\left(s_k\right)$. $V_\pi\left(s_k\right)$ also known as baseline functions, is estimated using the value network $v\left(s_k ; \boldsymbol{w}\right)$. The value network $v\left(s_k ; \boldsymbol{w}\right)$ is updated by minimizing the mean square error between $v\left(s_k ; \boldsymbol{w}\right)$ and the discount reward $U_{t}$.
\begin{align}
L(\mathbf{w})=\frac{1}{p-t} \sum_{k=t}^p\left[v\left(s_k ; \mathbf{w}\right)-u_t\right]^2
\end{align}

\subsubsection{PPO arithmetic}
One of the major drawbacks of policy gradient algorithms is their low sample efficiency. Trajectories sampled by the current policy $\pi\left(a_t \mid s_t ; \boldsymbol{\mu}\right)$ can only be used to update the current network, and once the network parameters are updated, new trajectories must be sampled. To overcome this drawback, literature \cite{TRPOn} proposed the Trust Region Policy Optimization algorithm (TRPO). TRPO introduces importance sampling, allowing the current policy network $\pi\left(a_t \mid s_t ; \boldsymbol{\mu}\right)$ to be updated based on experience trajectories sampled from the policy network at last episode $\pi\left(a_t \mid s_t ; \boldsymbol{\mu}_{\text {old }}\right)$. The TRPO algorithm updates the policy network by maximizing the surrogate objective function, which can be expressed as:
\begin{align}\label{TRPO}
\max L^{C P I}(\boldsymbol{\mu}, \mathbf{w})=\widehat{\mathbb{E}}_t\left[\rho_t(\boldsymbol{\mu}) \hat{A}_t\right]
\end{align}
where $\rho_t(\boldsymbol{\mu})\triangleq \pi\left(a_t \mid s_t ; \boldsymbol{\mu}\right) / \pi\left(a_t \mid s_t ; \boldsymbol{\mu}_{\text {old }}\right)$ denotes importance sampling weights and $\hat{A}(t)$ can expressed as:
\begin{align}
\hat{A}(t)=\sum_{l=0}^p(\gamma \lambda)^l \delta_{t+l}
\end{align}
where $\delta_t=r_t+\gamma v\left(s_{t+1} ; \boldsymbol{w}\right)-v\left(s_t ; \boldsymbol{w}\right)$. 
In order to effectively update the policy network parameter, it is only when the action distribution sampled from $\pi\left(a_t \mid s_t, \boldsymbol{\mu}\right)$ and the action distribution sampled from $\pi\left(a_t \mid s_t, \boldsymbol{\mu}_{\text {old }}\right)$ satisfy the trust region constraint \eqref{Trust} that $\pi\left(a_t \mid s_t, \boldsymbol{\mu}\right)$ can be updated using the experience trajectory sampled from $\pi\left(a_t \mid s_t, \boldsymbol{\mu}_{\text {old }}\right)$, 
\begin{align}\label{Trust}
\widehat{\mathbb{E}}_t\left[\mathrm{KL}\left[\pi\left(a_t \mid s_t ; \boldsymbol{\mu}_{\text {old }}\right), \pi\left(a_t \mid s_t ; \boldsymbol{\mu}\right)\right]\right] \leq \delta
\end{align}
Problem \eqref{TRPO} combined with constraint \eqref{Trust} is a constrained optimization problem, which must be solved using numerical optimization methods, such as conjugate gradient method and line search. To facilitate the computation, reference \cite{PPO} proposes the PPO algorithm. Problem \eqref{TRPO} combined with constraint \eqref{Trust} can be approximated by:
\begin{align}\label{actor_SGD}
&L^{C L I P}(\boldsymbol{\mu})= \nonumber\\
&\hat{\mathbb{E}}_t\left[\min \left(\rho_t(\boldsymbol{\mu}) \hat{A}_t, \operatorname{clip}\left(\rho_t(\boldsymbol{\mu}), 1-\varepsilon, 1+\varepsilon\right) \hat{A}(t)\right) \right.\nonumber\\
&\left.+c_2 S\left[\pi\left(a_t \mid s_t ; \boldsymbol{\mu}\right)\right]\right]
\end{align}
where $S$ denotes the entropy bonus, $c_{2}$ denotes the coefficients of entropy bonus and 'clip' refers to the clipping function, which can be expressed as:
\begin{align}
&\operatorname{clip}\left(\rho_t(\mu), 1-\varepsilon, 1+\varepsilon\right) = \nonumber\\
&\begin{cases}
1+\varepsilon, & \text{if } \rho_t(\mu)>1+\varepsilon \\
\rho_t(\mu), & \text{if } 1-\varepsilon \leq \rho_t(\mu) \leq 1+\varepsilon \\
1-\varepsilon, & \text{if } \rho_t(\mu)<1-\varepsilon
\end{cases}
\end{align}
The policy network is updated by the Stochastic Gradient Descent (SGD) algorithm, and the update formula can be expressed as:
\begin{align}\label{critic_SGD}
\boldsymbol{\mu} \longleftarrow \boldsymbol{\mu}--\frac{1}{B} \sum^B \nabla L^{C L I P}(\boldsymbol{\mu})
\end{align}
The parameters of the value network $v\left(s_t ; \boldsymbol{w}\right)$ are updated by minimizing the temporal-difference error \eqref{TD-error}.

\begin{align}\label{TD-error}
\frac{1}{2 B} \sum^B\left[r_t+\gamma \cdot v\left(s_{t+1} ; \boldsymbol{w}\right)-v\left(s_t ; \boldsymbol{w}\right)\right]^2
\end{align}

\subsection{MDP Formulation}
To solve the outer sub-problem using DRL, it is necessary to transform the outer sub-problem into MDP. The state space, action space and reward function are defined as follows:

\textbf{1) State Space:} The definition of the state space should contain as much environmental information as possible. For the outer optimization sub-problem, the environmental of its MDP includes Channel State Information(CSI) of all channel in the system (i.e., $\boldsymbol{h}^{\mathrm{RB}},\boldsymbol{h}_j^{\mathrm{U R}}, h_j^{\mathrm{U B}}, \boldsymbol{h}^{\mathrm{ES}}, \forall j \in \mathcal{J}$). Thus the state space of the MDP is defined as:
\begin{equation}
\mathcal{S}=\left\{\left\|\boldsymbol{h}^{\mathrm{RB}}\right\|,\left\|\boldsymbol{h}_j^{\mathrm{UR}}\right\|,\left|h_j^{\mathrm{UB}}\right|,\left\|\boldsymbol{h}^{\mathrm{ES}}\right\| \right\}, \forall j \in \mathcal{J}
\end{equation}
with the state space dimension $JN+N+J+1$

\textbf{2) Action Space:} The optimization variables of the outer sub-problem include the operating modes of the reflecting elements and the amplification factors. Therefore, the action space of the MDP consists of $\left\{\alpha_n, \beta_n, \rho_n \mid n \in \mathbb{N}\right\}$. To reduce the scale of the action space and accelerate training convergence, we discretizes $\rho_n $ into $\hat{\rho}_n$, where $\hat{\rho}_n \in\left\{10 \times\left\lceil\frac{\rho_n}{10}\right\rceil \mid \rho_n \in\left[0, \rho_{\max }\right], n \in \mathbb{N}\right\}$ and $\lceil\cdot\rceil$ is the floor function operator. Correspondingly, the action space of the MDP can be defined as:
\begin{align}\label{action}
\mathcal{A}=\left\{\alpha_n, \beta_n, \hat{\rho}_n \mid n \in \mathbb{N}\right\}
\end{align}
When $\rho_{\max }=100$, the size of the discrete action space defined by Equation \eqref{action} is $10^N \times 2^N \times 2^N$, which increases exponentially with the number of reflecting elements. To solve the problem of the action space being too large, leading to difficulties in training convergence, this paper employs the action composition technique from literature \cite{Gym-µRTS}. By applying a logarithmic transformation to the gradients of the policy network, the size of the discrete action space is reduced from the original $10^N \times 2^N \times 2^N$ to $(10+2+2) \times N$. This approach can accelerate the speed of training convergence.

\textbf{3) reard function:} It is noted that problem P2 is not always feasible given $\left\{\alpha_n, \beta_n, \rho_n \mid n \in \mathbb{N}\right\}$. For example, when $\beta_{\mathrm{n}}=0, n \in N$ and $Q_{min}$ is relatively large, constraint \eqref{Q_min} cannot be satisfied regardless of the values of the optimization variables $\left\{E_{i, j}, t_{i, j}\right\}$. Therefore, in this case, there is no feasible solution for problem P2. Consequently, we defines the immediate reward as:
\begin{equation}
\text { reward }= \begin{cases}-F^*(\boldsymbol{\alpha}, \boldsymbol{\beta}, \boldsymbol{\rho}), & \text { if problem } \mathrm{P} 2 \text { is solvable } \\ -\eta, & \text { otherwise }\end{cases}
\end{equation}
where $F^*(\boldsymbol{\alpha}, \boldsymbol{\beta}, \boldsymbol{\rho})$ is the optimal value of the objective function of problem P2, and $\eta$ is a sufficiently large positive number, indicating that the reward obtained when problem P2 has no feasible solution is very small.
\subsection{Algorithm Implementation}
Based on the aforementioned content, it can be summarized that the specific process for solving problem P1 using the PPO algorithm is as shown in Algorithm 1.
\begin{algorithm}[H]
\caption{PPO-Convex}
\begin{algorithmic}[1]
    \State Initialize actor network $\boldsymbol{\mu}$, critic network $\mathbf{w}$
    \For {each episode}:
        \State Reset the environment and observe initial state $s_{t}$, 
        \Statex \hspace{1cm}$t=0$.
        \For{each step}:
        \State Normalization $s_{t}$ and sample action $a_{t}$ based on 
        \Statex \hspace{1cm}
        $\pi\left(a_t \mid s_t ; \boldsymbol{\mu}\right)$.
        \State Solve P2 and get reward $r_{t}$
        \State Observe next state $s_{t+1}$
        \State Save $(s_{t},a_{t},r_{t},s_{t+1})$ into buffers, $\mathrm{s}_{\mathrm{t}} \leftarrow \mathrm{s}_{\mathrm{t}+1}$  
        \EndFor
    \For {each update}:
        \State Sample mini batch from the buffer
        \State Update actor network based on \eqref{actor_SGD} via SGD
        \State Update critic network based on \eqref{critic_SGD} via SGD
    \EndFor
    \State Clear the buffer
    \EndFor
\end{algorithmic}
\end{algorithm}
At the beginning, Initialize the actor-network parameter $\boldsymbol{\mu}$ and the critic network parameter $\mathbf{w}$, and initialize the environment settings (line1-line3). Subsequently, the Z-Score Normalization method is applied to standardize the state $s_{t}$. Then, the actor network takes the standardized state $s_{t}$ as input and outputs the probability distribution of action $a_{t}$, $\pi\left(a_i \mid s_i ; \boldsymbol{\mu}\right)$. Based on this probability distribution $\pi\left(a_i \mid s_i ; \boldsymbol{\mu}\right)$, Sample actions $\alpha_{n}$, $\beta_{n}$, and $\rho_n$ (Line 4-5). After determining $\alpha_{n}$, $\beta_{n}$, and $\rho_n$, Solve problem P2 to obtain the reward $r_{t}$ (Line 6). Next, Update the state of the wireless channel from $s_{t}$ to $s_{t+1}$, store the transition $\left(s_t, a_t, r_t, s_{t+1}\right)$ in the buffers (Line 7-Line 8). We repeat Line 5-Line 8 until the for loop ends. Then, Sample a mini-batch transition $\left(s_t, a_t, r_t, s_{t+1}\right)$ from the buffers and update the policy network and critic network using the Stochastic Gradient Descent (SGD) method, repeating (Line 12-Line 14) until the maximum number of updates is reached. Repeat the entire process (Line 3-Line 14) until the maximum number of episode is reached. Figure 3 further outlines the process of the algorithm.
\begin{figure}[ht]
\centering
\includegraphics[width=3.3in]{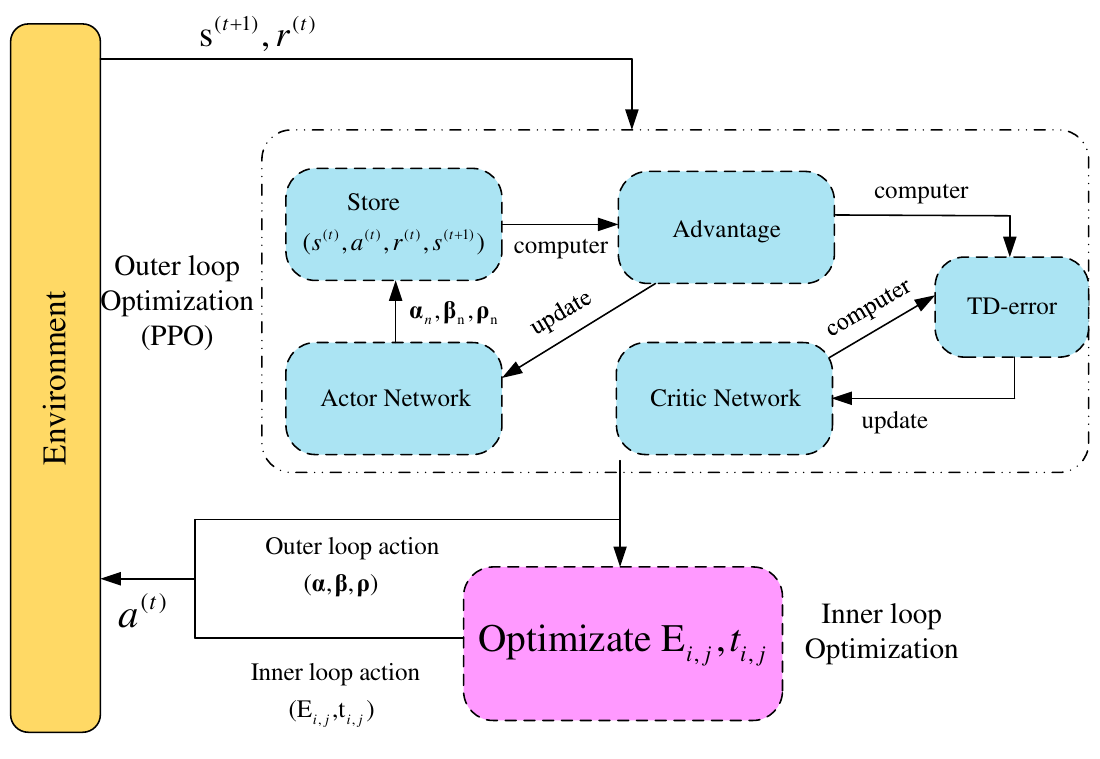}
\centering
\caption{The structure of time frame}
\end{figure}
% Numbered list
% Use the style of numbering in square brackets.
% If nothing is used, the default style will be taken.
%\begin{enumerate}[a)]
%\item 
%\item 
%\item 
%\end{enumerate}  

% Unnumbered list
%\begin{itemize}
%\item 
%\item 
%\item 
%\end{itemize}  

% Description list
%\begin{description}
%\item[]
%\item[] 
%\item[] 
%\end{description}  

% Uncomment and use as the case may be
%\begin{theorem} 
%\end{theorem}

% Uncomment and use as the case may be
%\begin{lemma} 
%\end{lemma}

%% The Appendices part is started with the command \appendix;
%% appendix sections are then done as normal sections
%% \appendix
\section{Simulation result}
This section compares the proposed algorithms to baseline algorithms in order to evaluate the effectiveness of the proposed algorithms. The baseline algorithms are defined as:
\begin{itemize}
  \item \textbf{PPO-Active-Passive-RIS:} The amplification factors of active element and the operation mode of each reflecting units (eg. active mode and passive mode) are solved in outer loop by PPO algorithm, but the transmit power and time allocation are determine by convex optimization methods in inner loop.
  \item \textbf{PPO-Active-RIS:} The amplification factors of active RIS are solved in outer loop by PPO algorithm, but the transmit power and time allocation are determine by convex optimization methods in inner loop.
  \item \textbf{Passive RIS:} All reflecting elements are operation in passive mode.
  \item \textbf{Without RIS:} All reflecting elements are operating in idle mode.
\end{itemize}
The xy-plane coordinates as shown in Fig 4. In the simulation, it is assumed that the user terminals are uniformly distributed around a circle with a radius of 0.5m and centered at (0, 0, 0). The BS is located at (20m, 0m, 0), while RIS and EHS are located at (5m, 3m, 0m) and (5m, 3m, 5m) respectively, with the ES at (5m, -2m, 5m). The path loss exponents for the BS-AP, BS-RIS, and RIS-AP links are 3.2, 2.2, and 2.2 respectively, and the path losses at a reference distance of 1 m are -30 dB, -20 dB, and -20 dB respectively.
 \begin{figure}[h]
\centering
\includegraphics[width=3.3in]{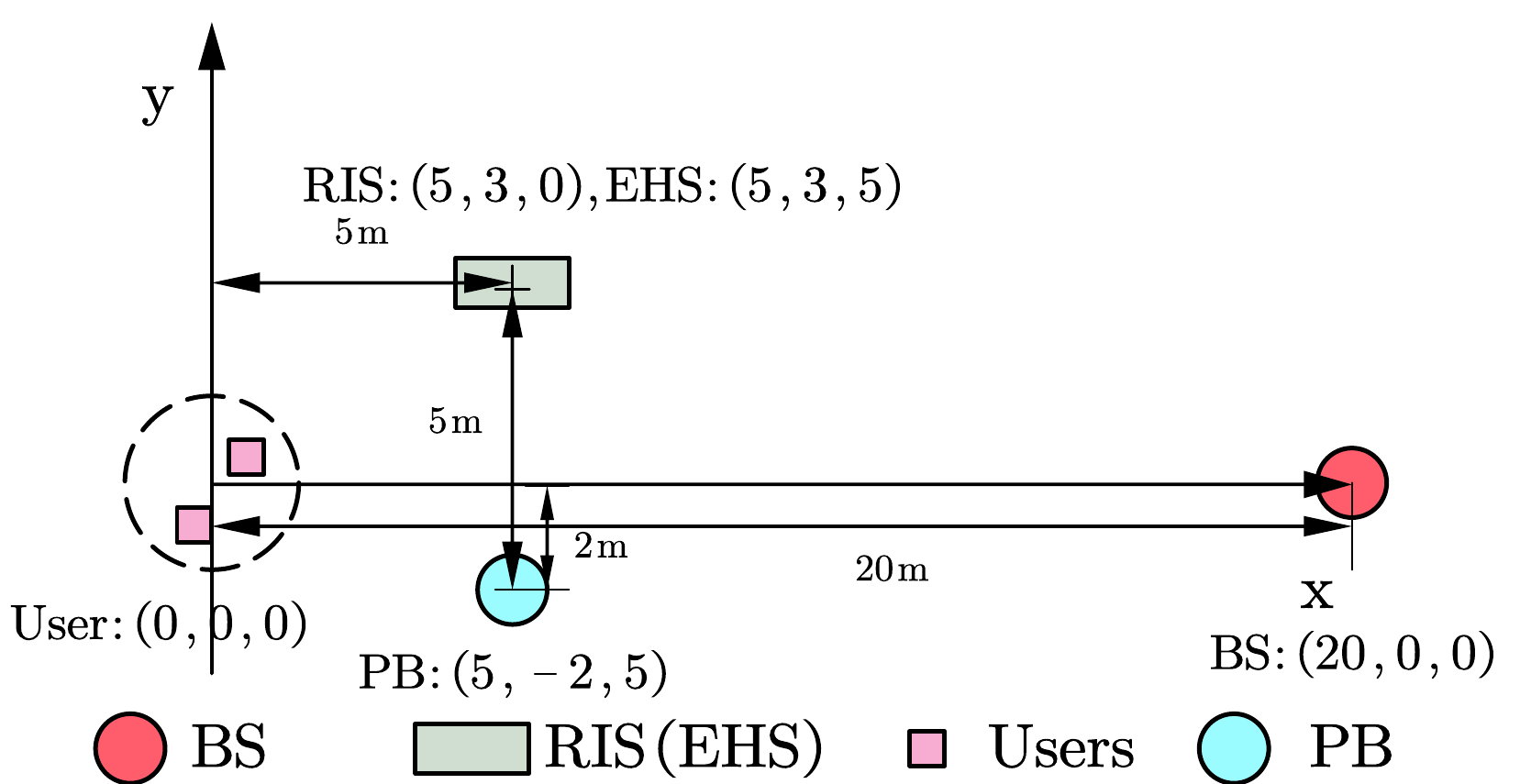}
\centering
\caption{The structure of time frame}
\end{figure}
The other simulation parameters of the system are set as follows:$P_{s} = 38$dBm, $Q_{min}=5$bits/Hz/s, $\sigma_{B}^{2}=-80$dBm, $\sigma_{F}^{2}= -70$dBm \cite{Active_Passive}, $P_{C}=-10$dBm, $P_{DC}=-5$dBm \cite{Active_first}, $P_{b}= 2.1 \mu \mathrm{W}$ \cite{Robust-and-Secure}, $M=N=20$, $J=2$. 

The parameter settings for the PPO algorithm are as follows:
\begin{table}[pos=H]
\caption{\centering Hyper Parameters Settings Of PPO Algorithm}
\begin{tabular}{|c|c|}
\hline
\textbf{Description} & \textbf{Setting} \\ \hline
The learning rate of the critic network & 0.0001 \\ \hline
The learning rate of the actor network & 0.0001 \\ \hline
Clipping parameter & 0.2 \\ \hline
Discount factor $\epsilon$ & 0\\ \hline
The number of episodes & 50000 \\ \hline
The number of steps & 512 \\ \hline
The number of mini-batches size & 128 \\ \hline
The numbers of neurons for hidden layers $\epsilon$ & 1024\\ \hline
The numbers of hidden layers & 1 \\ \hline
The coefficients of entropy bonus & 0.0001 \\ \hline
\end{tabular}
\end{table}
\subsection{The Convergence of PPO-Convex Arithmetic}

We simulated the convergence performance of PPO-Convex under different user scenarios, as shown in Fig 5. It can be observed that with the increase of episodes, min-max energy consumption gradually decreases and eventually converges. Furthermore, the min-max energy consumption is increasing with the number of users. This is because adding more users to the system implies that less time can be allocated for each user's transmission, which requires users to increase their transmit power to satisfy the minimum data rate requirement of the user.

\begin{figure}[ht]
\centering
\includegraphics[width=3.3in]{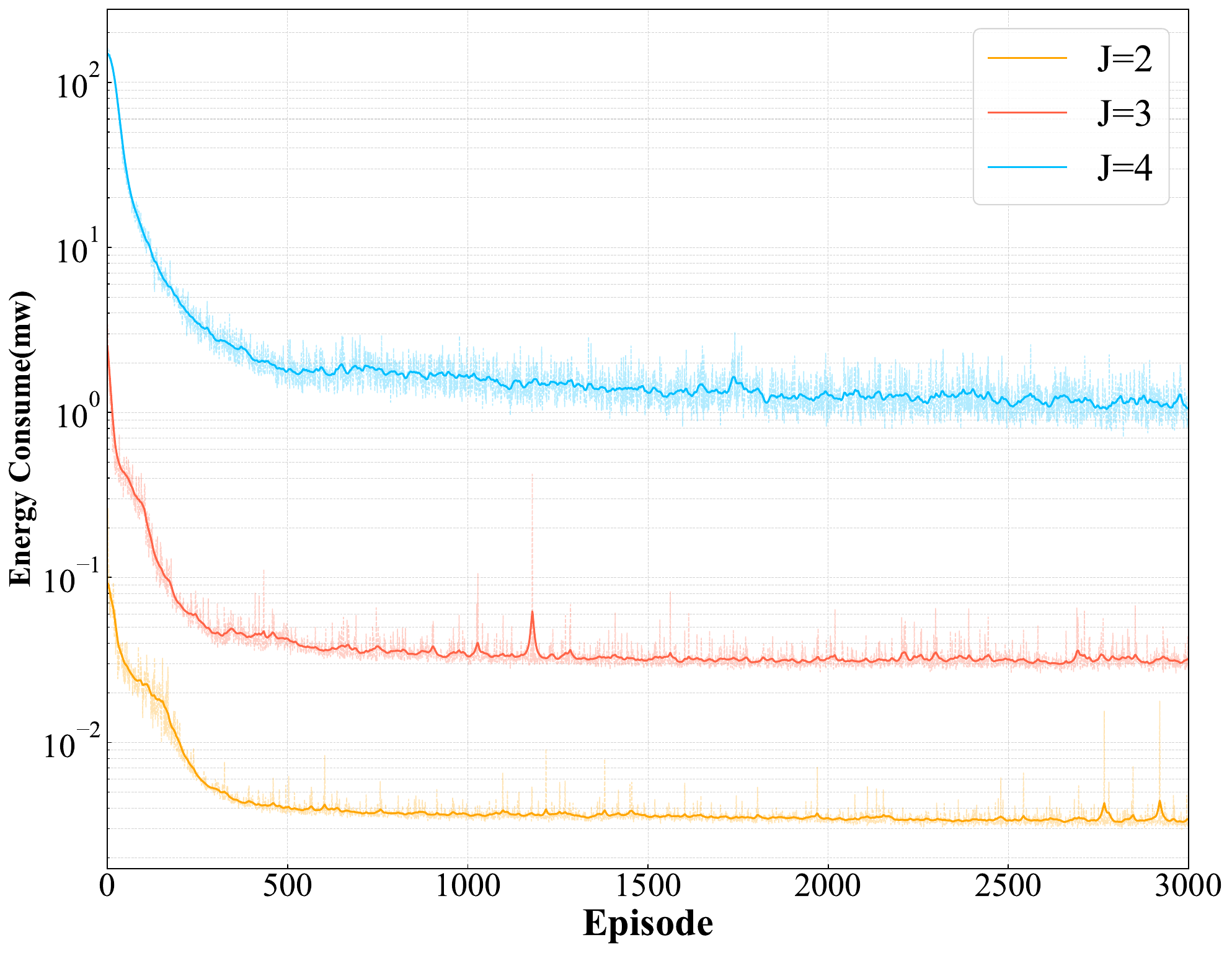}
\centering
\caption{The convergence of rewards for the different amount of users}
\end{figure}

Fig 6. shows the convergence performance of the PPO-Convex algorithm compared to the traditional model-free PPO algorithm in which all the optimizing variables are solved by the PPO method. From Fig 6., it can be observed that the PPO-Convex algorithm significantly reduces the maximum energy consumption of users compared to the standard PPO algorithm. This is because the PPO-Convex algorithm decomposes the problem into inner and outer problems and only focuses on learning the outer problem, reducing the action space for the algorithm's learning process.

\begin{figure}[ht]
\centering
\includegraphics[width=3.3in]{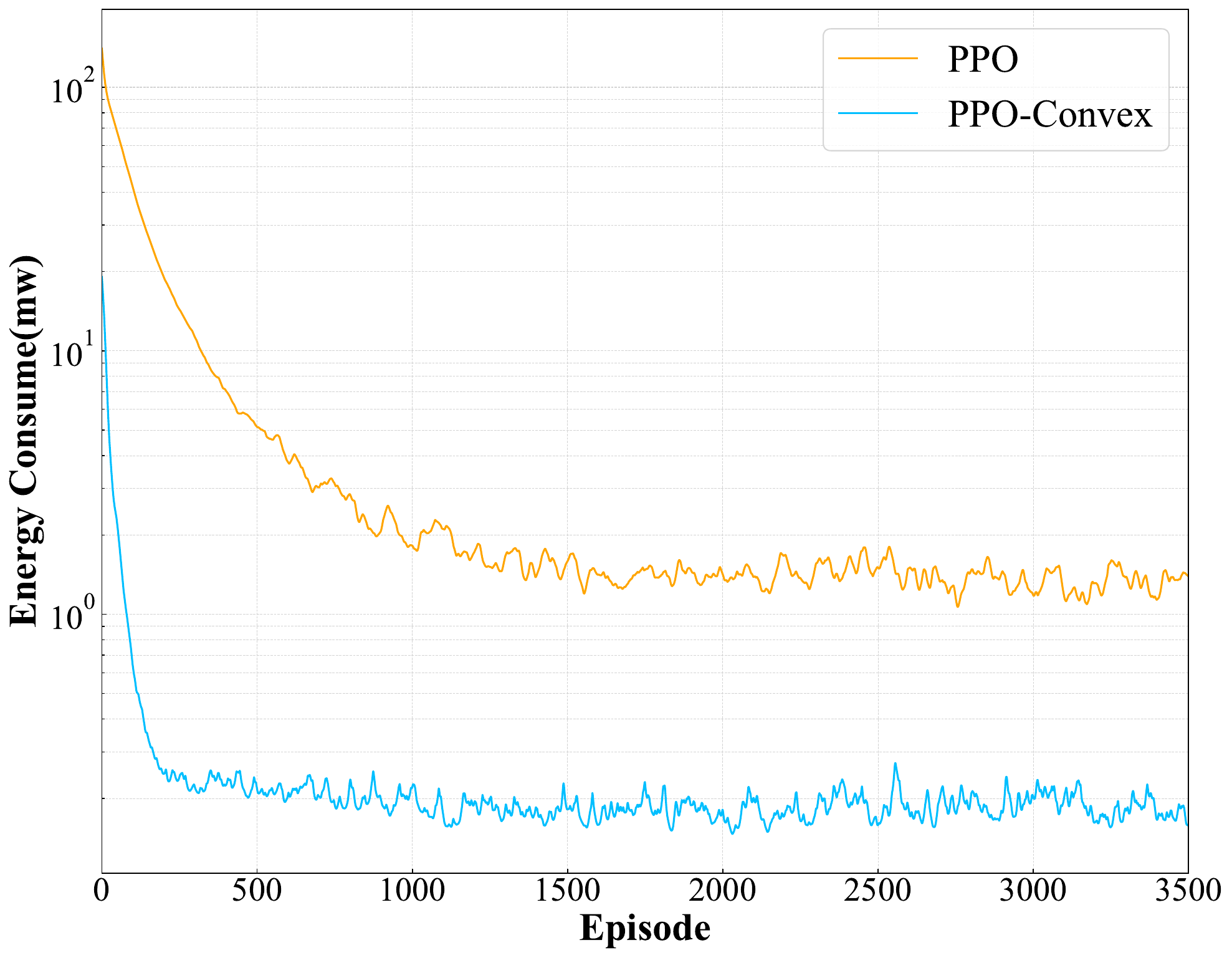}
\centering
\caption{PPO vs PPO-Convex: Convergence Analysis}
\end{figure}
\subsection{The influence of the ES’s transmit power on
min-max energy consumption}
Fig 7 illustrates the min-max energy consumption and the ratio of operation mode versus different transmit power of ES under different algorithms. From Fig 7(a), it can be observed that min-max energy first decreases and then gradually stabilizes with the increasing transmit power of ES. This is because the larger the transmit power of ES, the more energy is harvested by EHS, which undoubtedly increases the during of the RIS-assisted transmissions phase (i.e. $t_{2}$) and the amplification factor of active RIS until the $t_{2}$ and amplification factor reaches the maximum value. From Fig 7(b), it can be observed that the ratio of the active element is increasing and the ratio of the idle element and passive element is decreasing with the increasing transmit power of ES. This is because increasing the transmit power of ES leads to a significant rise in energy harvested by EHS, driving hybrid RIS to operate more elements in active mode. Furthermore, when the transmit power of ES is large ($45$dbm,$50$dbm), hybrid RIS has the same performance as the active-passive RIS and active RIS due to all reflecting elements of hybrid RIS and passive-active RIS operating in active mode when energy harvested by EHS is sufficient, as shown in  Fig 7(b). Finally, it's worth mentioning that active-passive RIS and passive RIS have the same performance when the transmit power of ES is equal to $30$dbm and $35$dbm since active-passive RIS tends to switch all reflecting elements to the passive mode when the transmit power of ES is small, as shown in Fig 7(a).
\begin{figure}[ht]
    \centering
    \begin{subfigure}[b]{0.48\textwidth}
        \includegraphics[width=\textwidth]{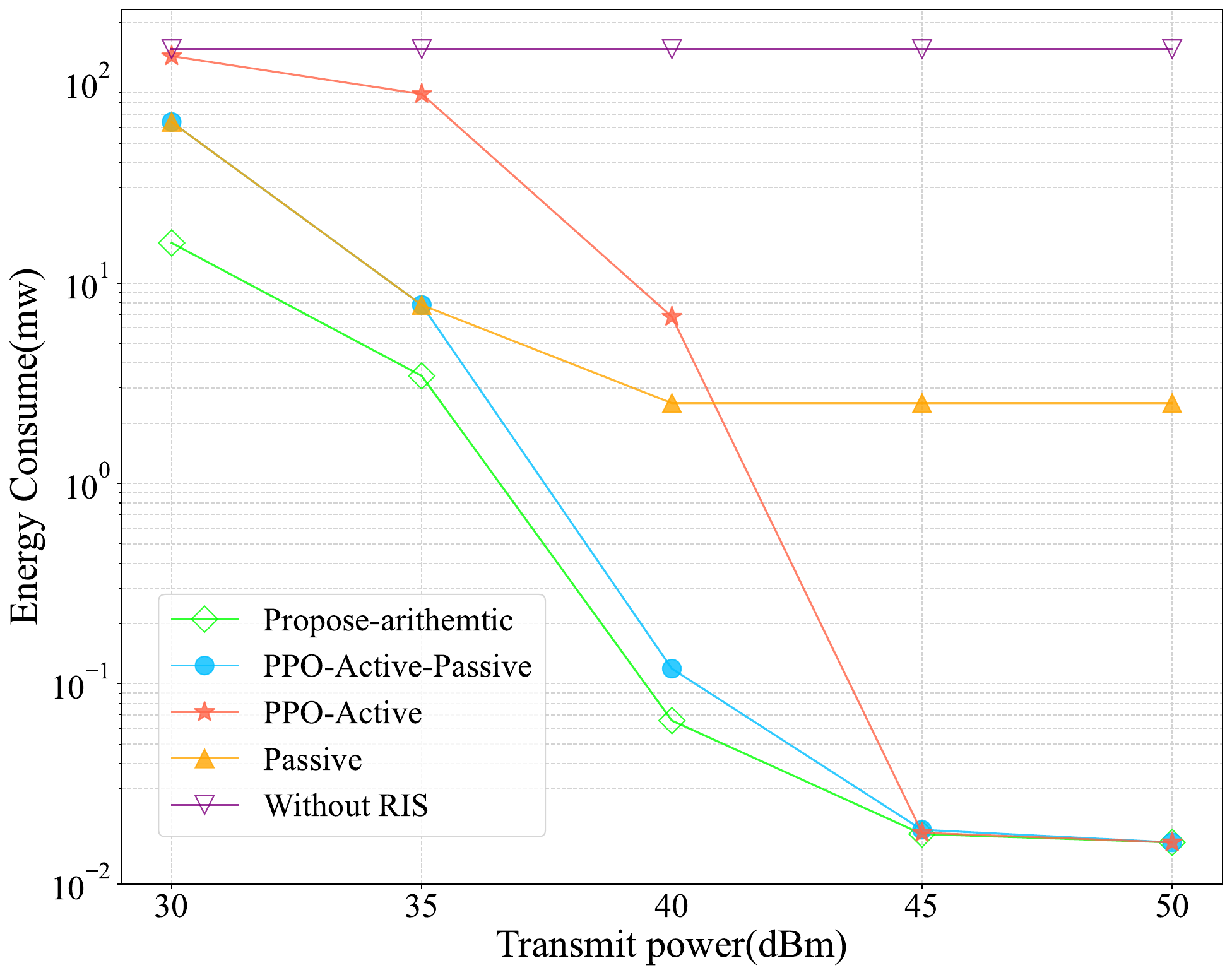}
        \caption{Min-max energy versus the transmit power of ES}
    \end{subfigure}
    \vspace{0cm} % 这里添加垂直空间，1cm是一个例子，您可以根据需要调整
    \begin{subfigure}[b]{0.48\textwidth}
        \includegraphics[width=\textwidth]{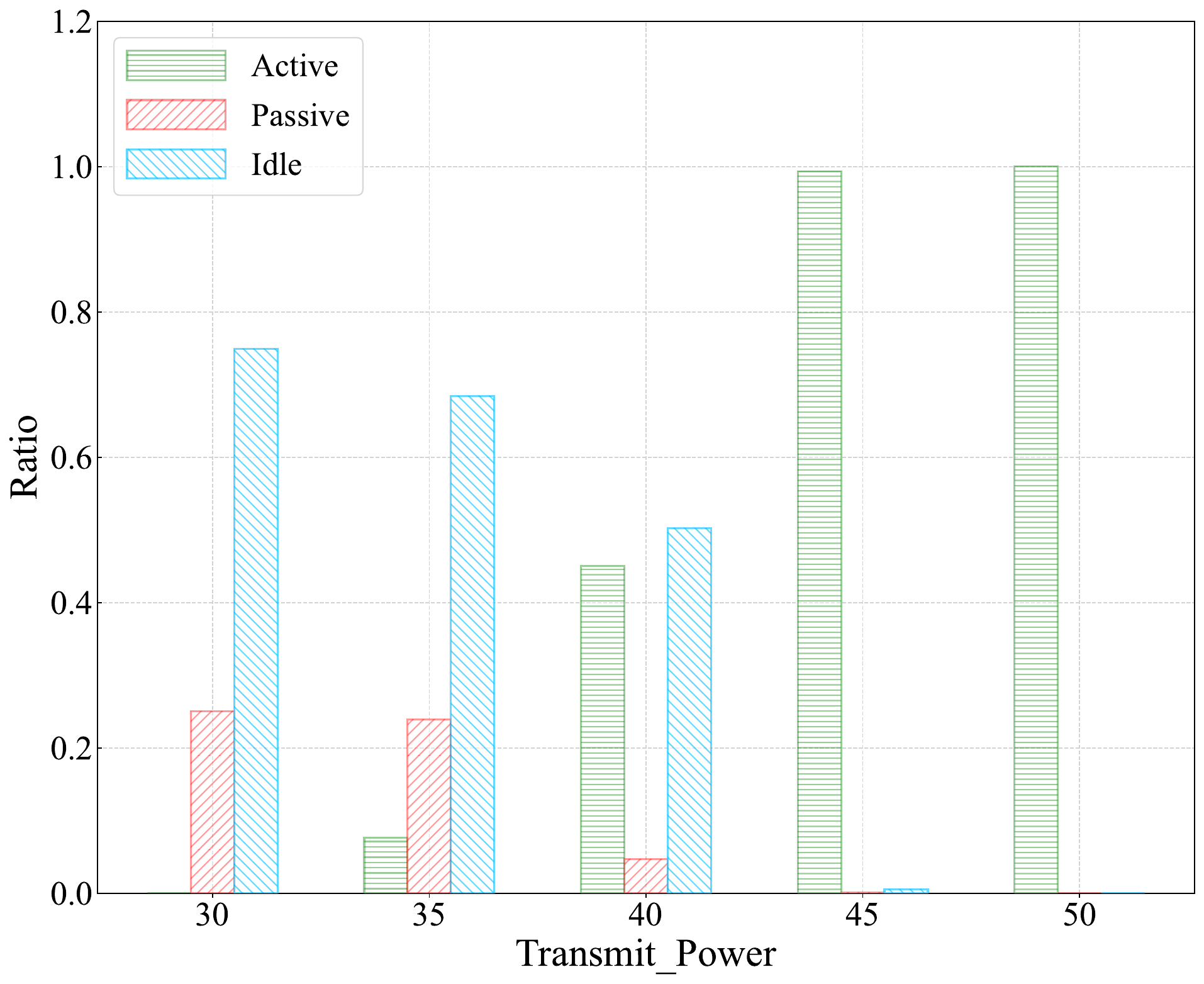}
        \caption{The ratio of operation mode versus the transmit power of ES}
    \end{subfigure}
    \caption{Min-max energy and the ratio of operation mode versus the transmit power of ES}
    %\vspace{-0.5cm}
\end{figure}

Fig 8 illustrates $t_{2}$ and the average amplification factor of active element versus different transmit power of ES under PPO-Active-Passive arithmetic and proposes arithmetic. They reveal the reason for the higher performance gains of proposed self-powered RIS schemes compared to other self-powered RIS schemes when the transmit power equals $30$dbm, $35$dbm, and $40$dbm, respectively. To begin with, since the logarithmic function is a decreasing growth rate function, extending $t_{2}$ rather than improving channel capacity, is more likely to enable users to meet QoS constraints. Hybrid RIS can further reduce the min-max energy consumption by operating some reflecting elements in idle mode to save energy, thereby extending $t_{2}$, as shown in Fig. 7(b) and Fig. 8(a). Furthermore, the hybrid RIS can operate some elements in idle mode thereby concentrating energy on a limited number of active elements. This approach ensures that the amplification factor of the active elements remains high to mitigate the “multiplicative fading” effect, effectively, as shown in Fig. 7(b) and Fig. 8(b). It is also worth mentioning that, the EHS harvests minimal energy when the transmit power equals $30$dbm. As a result, no elements operate in active mode, leading to an amplification factor of zero, as shown in Fig. 8(b). 

\begin{figure}[ht]
    \centering
    \begin{subfigure}[b]{0.48\textwidth}
        \includegraphics[width=\textwidth]{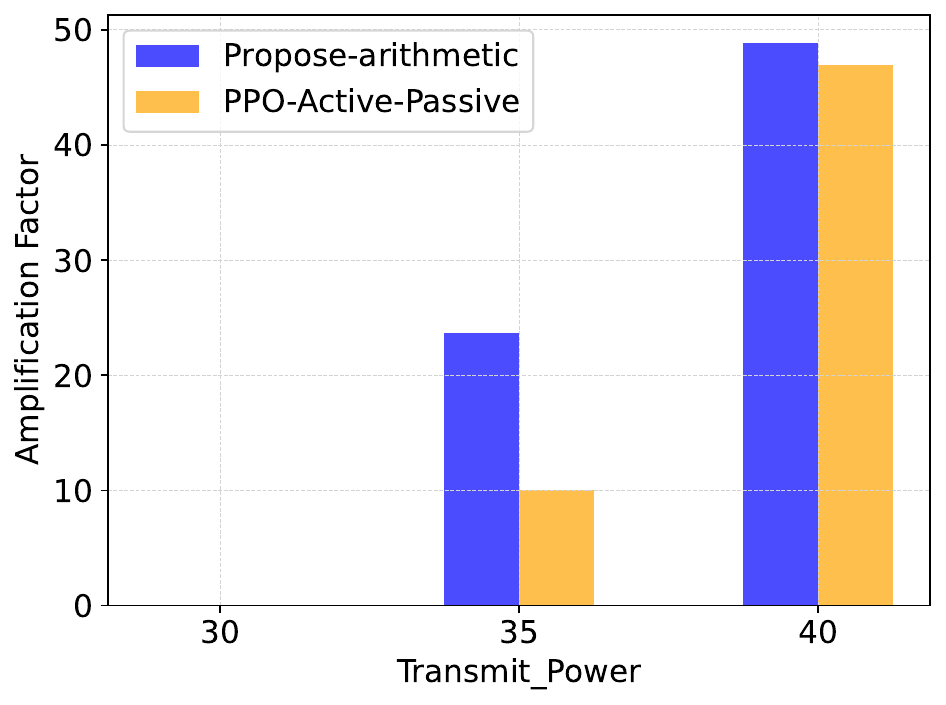}
        \caption{Amplification factor versus the transmit power of ES}
    \end{subfigure}
    \vspace{0cm} % 这里添加垂直空间，1cm是一个例子，您可以根据需要调整
    \begin{subfigure}[b]{0.48\textwidth}
        \includegraphics[width=\textwidth]{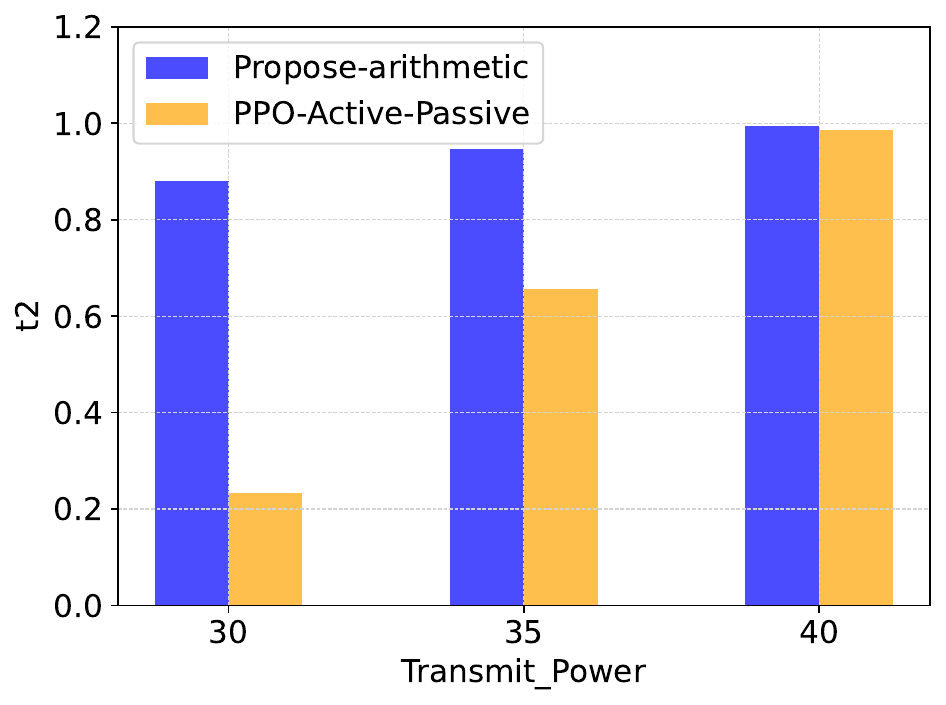}
        \caption{$t_{2}$ versus the transmit power of ES}
    \end{subfigure}
    \caption{Amplification factor and $t_{2}$ versus the transmit power of ES}
\end{figure}

\begin{figure}[ht]
    \centering
    \begin{subfigure}[b]{0.48\textwidth}
        \includegraphics[width=\textwidth]{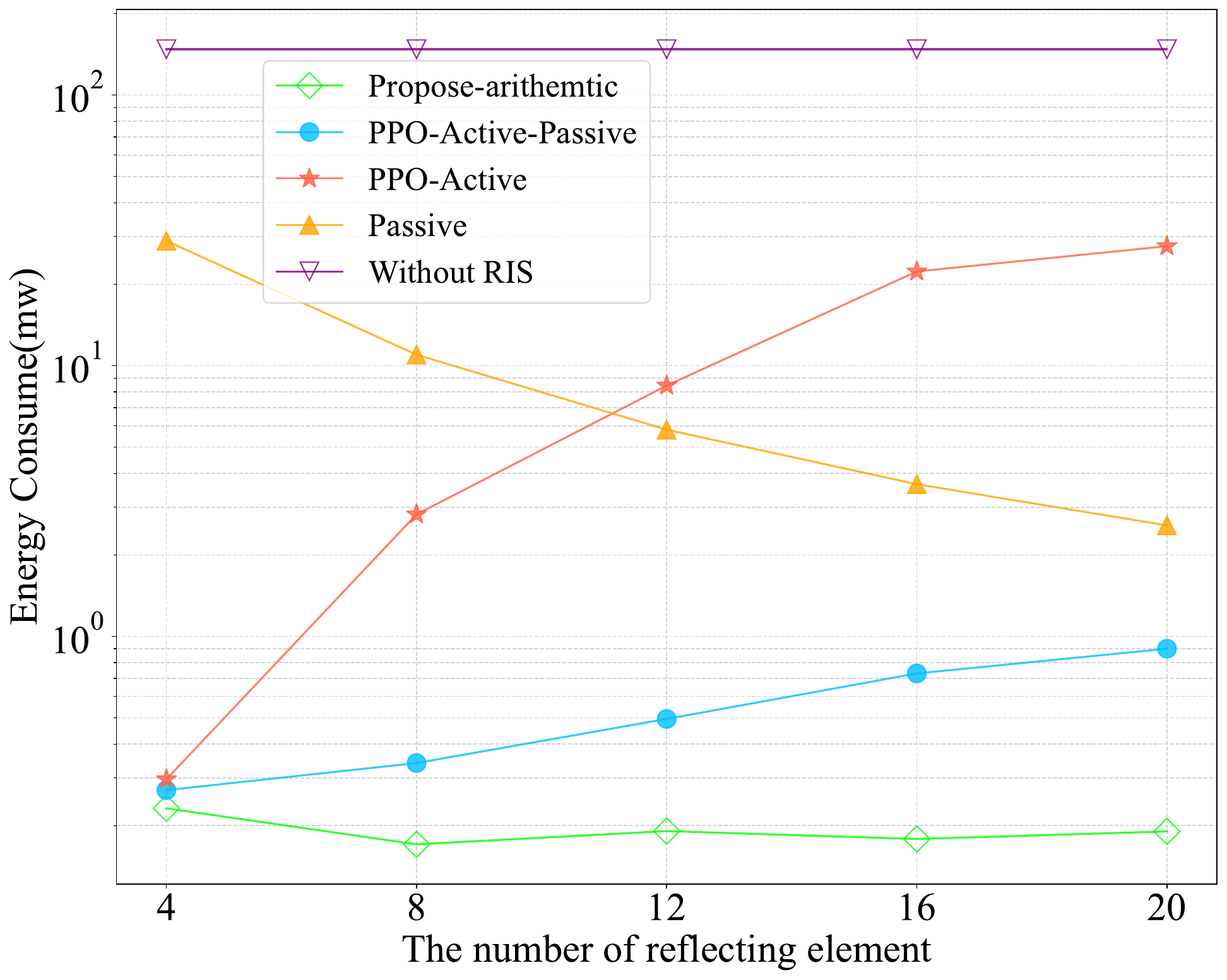}
        \caption{Min-max energy versus the number of elements of RIS}
    \end{subfigure}
    \vspace{0cm} % 这里添加垂直空间，1cm是一个例子，您可以根据需要调整
    \begin{subfigure}[b]{0.45\textwidth}
        \includegraphics[width=\textwidth]{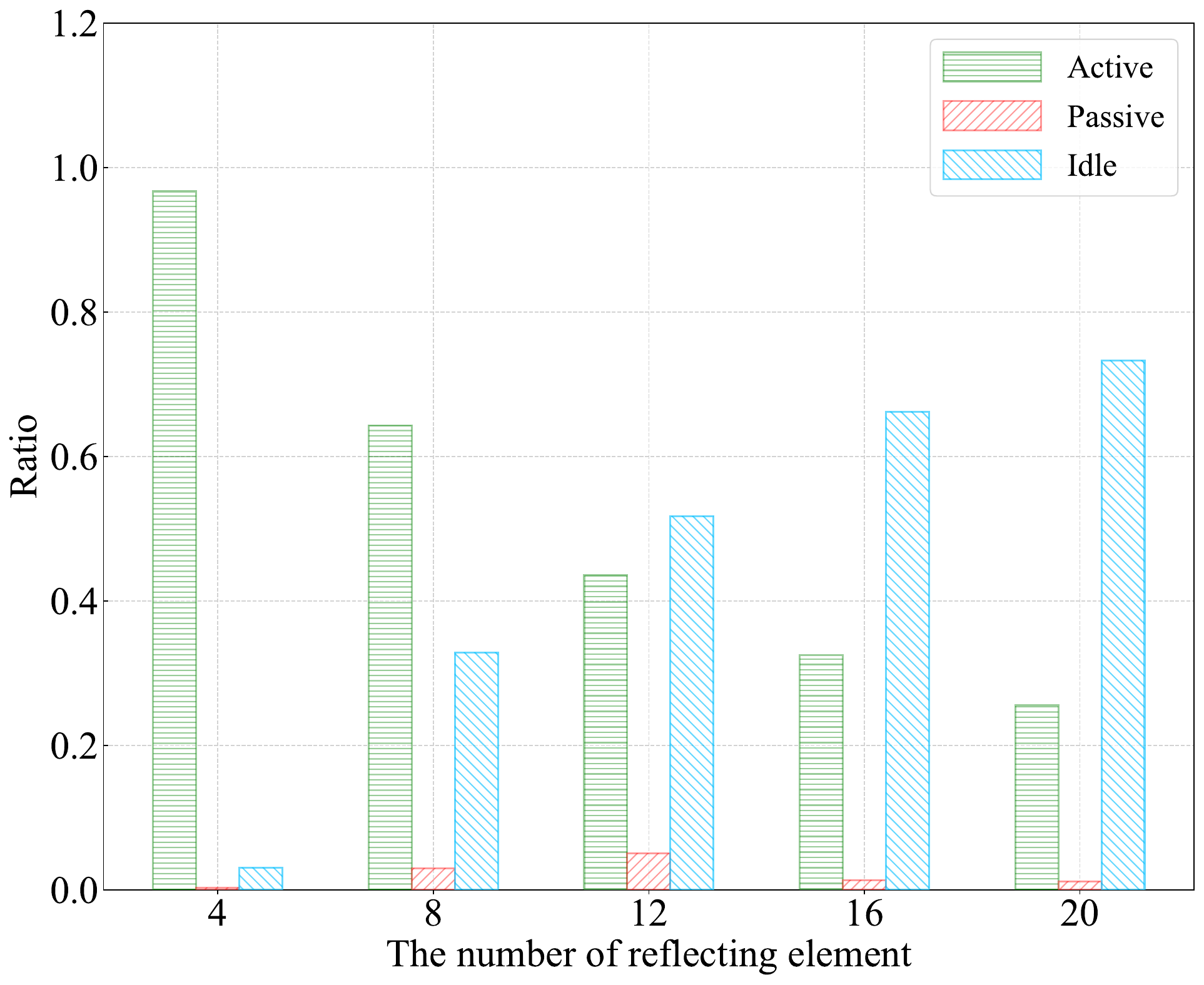}
        \caption{The ratio of operation mode versus the number of elements of RIS}
    \end{subfigure}
    \caption{Min-max energy and the ratio of operation mode versus the number of elements of RIS}
\end{figure}

Fig.9 shows the min-max energy consumption and the ratio of operation mode versus the number of reflecting elements under different algorithms. From Fig.9(a), it can be seen that with the increasing reflecting element, the min-max energy consumption of the system under the PPO-active-algorithm and PPO-active-passive-algorithm increases. This is because increasing the number of reflecting elements also increases the circuit's power consumption, which results in less time on signal reflection and a low amplification factor of the active element. However, we can obtain quite different conclusions for the passive RIS scheme. This is because passive RIS has lower power consumption. So increasing the number of passive RIS elements does not reduce $t_{2}$. On the contrary, it can provide additional transmission links for signal transmission. Finally, our
proposed RIS architecture can greatly minimize the min-max energy consumption of users as compared with the baseline schemes since the proposed hybrid RIS can
operate in some reflecting element in idle mode, which
remains the number of active, passive elements, $t_{2}$, and application factor of active element at an optimal level, as shown in Fig.9(b).

\begin{figure}[ht]
    \centering
    \begin{subfigure}[b]{0.48\textwidth}
        \includegraphics[width=\textwidth]{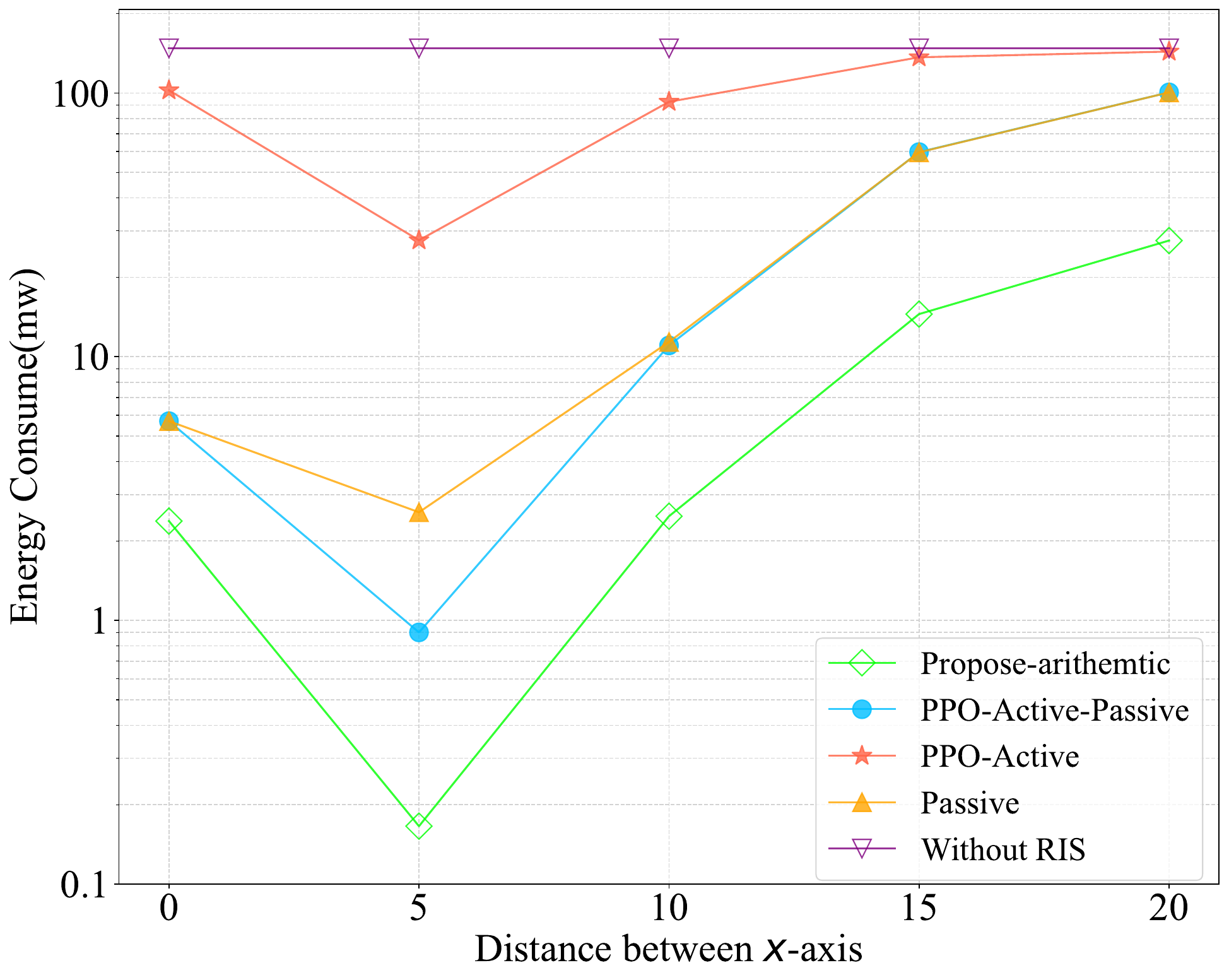}
        \caption{Min-max energy versus distance between the RIS and EHS to the x-axis}
    \end{subfigure}
    \vspace{0cm}  % 这里添加垂直空间，1cm是一个例子，您可以根据需要调整
    \begin{subfigure}[b]{0.48\textwidth}
        \includegraphics[width=\textwidth]{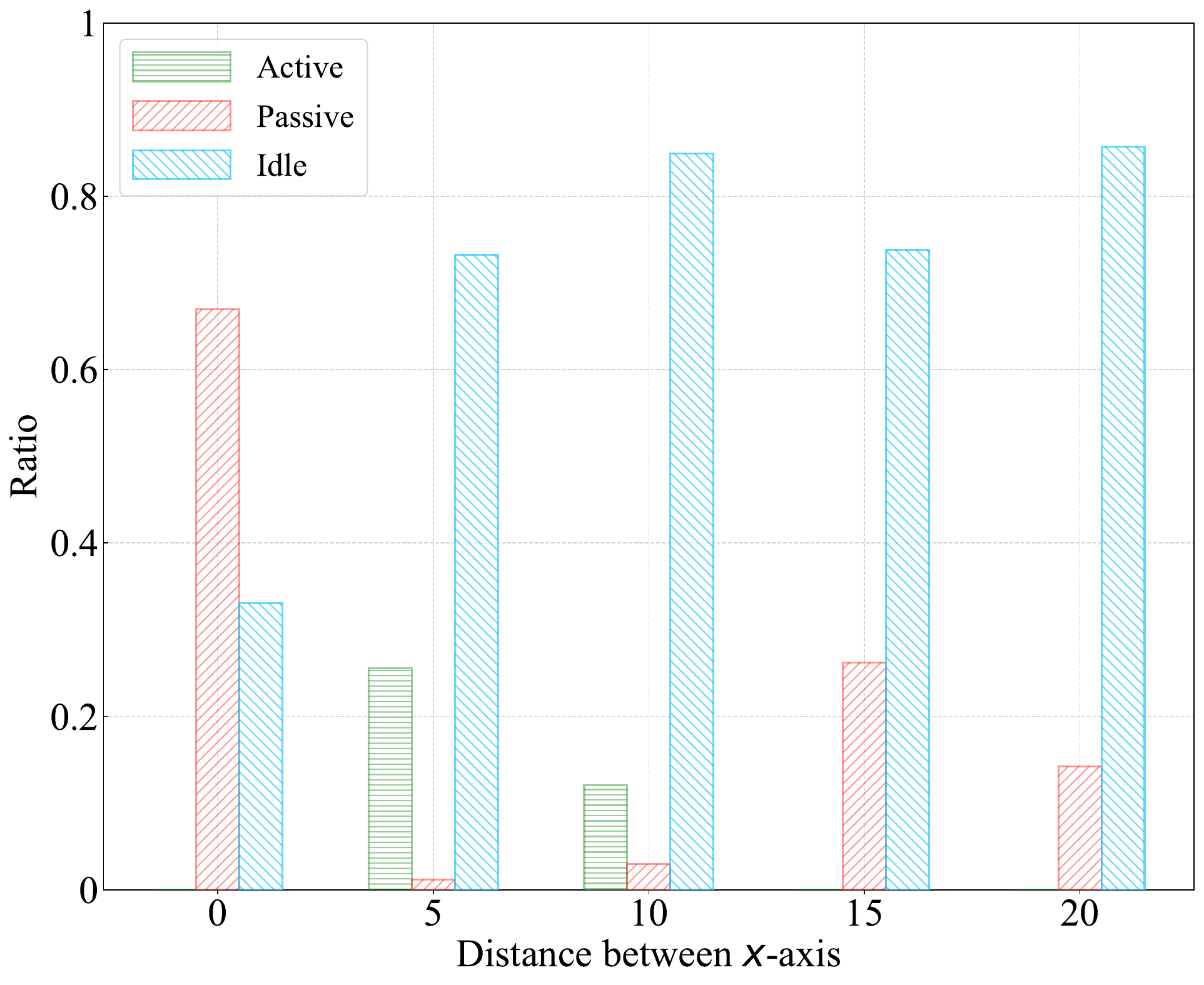}
        \caption{The ratio of operation mode versus distance between the RIS and EHS to the x-axis}
    \end{subfigure}
    \caption{Min-max energy and the ratio of operation mode versus distance between the RIS and EHS to the x-axis}
\end{figure}
Fig.10 shows the min-max energy consumption and the ratio of operation mode versus the distance between RIS and EHS to the x-axis. From Fig.10(a), it can be observed that as the distance between the RIS and EHS to the x-axis increases, the min-max energy consumption first reduces and then increases. The reason is illustrated as follows. When the RIS and EHS are getting close to ES, the energy harvested by EHS increases. Thus, the RIS can switch more reflecting elements to active mode, which improves channel power gains, leading to a reduction in the transmit power of the user, as shown in Fig,10(a) and Fig,10(b). However, as the RIS and EHS move away from the ES, the energy harvested by the EHS decreases, leading to fewer elements operating in active mode. Thus, the user has to increase the transmit power to meet the QoS constraint, as shown in Fig,10(a) and Fig,10(b).
It is worth mentioning that the energy harvested by EHS is identical when the RIS and EHS are located in $0$m and $10$m, respectively.When EHS and RIS are located in $0$m, the number of active elements is zeros. In contrast, when EHS and RIS are located at $10$m, the number of active elements is approximately $2$. It is because increasing the number of active elements can provide amplification gain for the signal but also increases the energy consumption, which reduces the $t_{2}$. If the improved amplification gain can compensate for the reduction of $t_{2}$, the min-max energy consumption can decrease; otherwise, the min-max energy consumption increases. When the EHS and RIS are closer to the transmitter, the signal power received at the RIS is higher. As a result, the amplification of the signal by the RIS requires greater energy consumption, leading to a reduced amplification factor and less time on signal reflection so that active mode isn't selected as the optimal operation mode when the EHS and RIS located in $0$m.

\begin{figure}[ht]
    \centering
    \begin{subfigure}[b]{0.48\textwidth}
        \includegraphics[width=\textwidth]{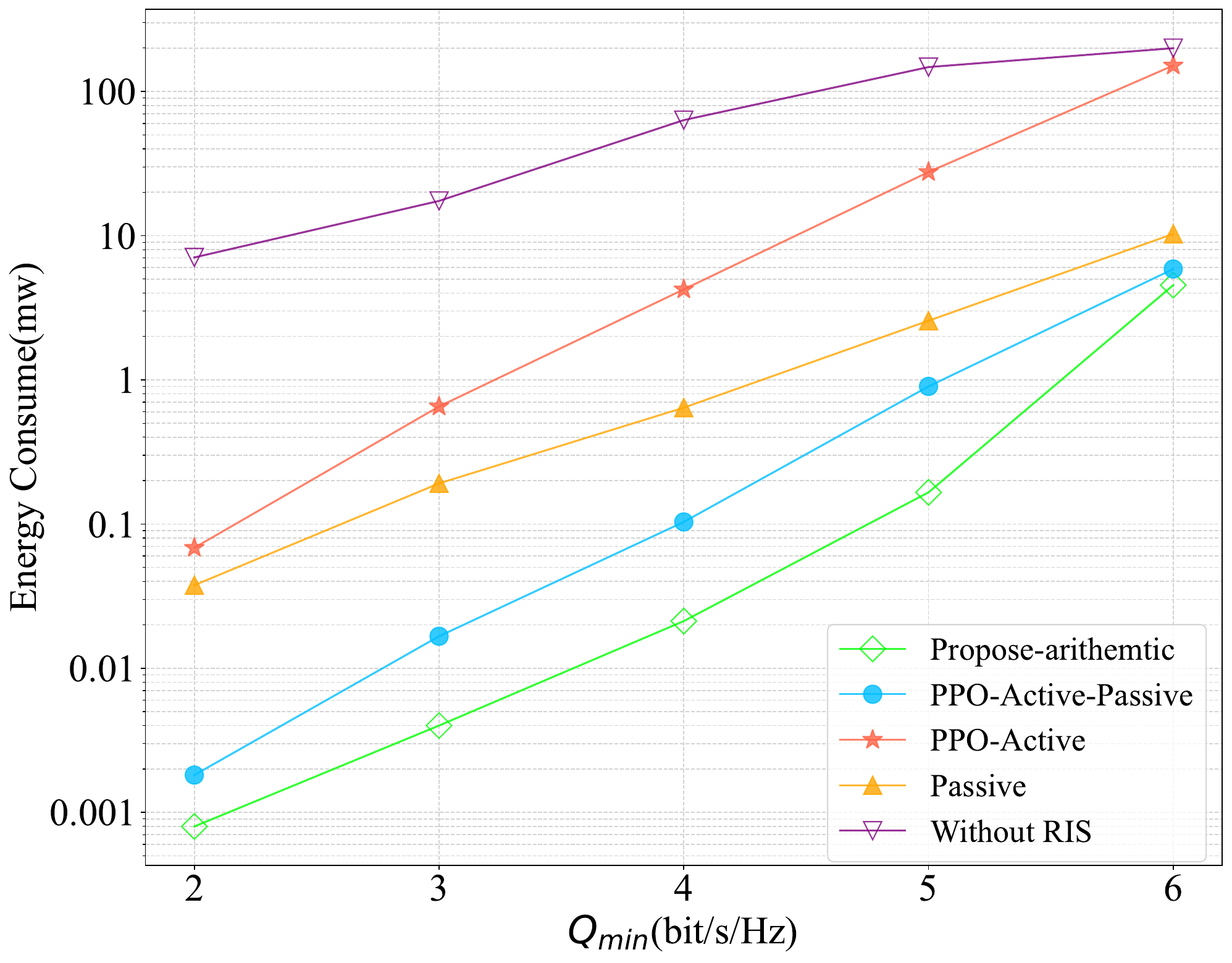}
        \caption{Min-max energy versus $Q_{min}$}
    \end{subfigure}
    \vspace{0cm} % 这里添加垂直空间，1cm是一个例子，您可以根据需要调整
    \begin{subfigure}[b]{0.48\textwidth}
        \includegraphics[width=\textwidth]{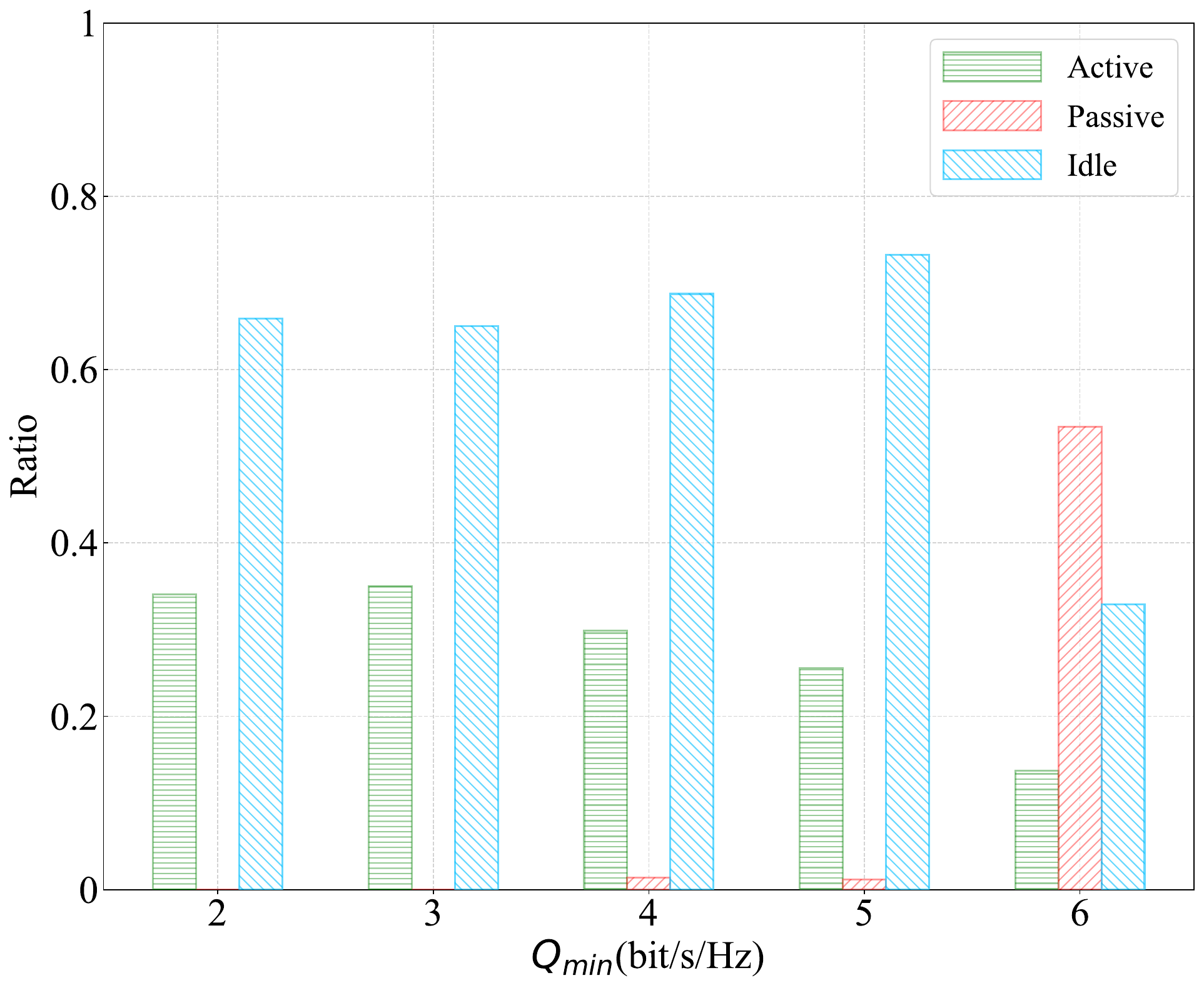}
        \caption{The ratio of operation mode versus $Q_{min}$}
    \end{subfigure}
    \caption{Min-max energy and the ratio of operation mode versus $Q_{min}$}
    %\vspace{-0.5cm}
\end{figure}

Fig 11 shows the min-max energy consumption and the ratio of operation mode versus the minimum transmission rate under different algorithms. From Fig. 11(a), it can be seen that the increasing the minimum transmission rate, the min-max energy consumption under all algorithms increases. This is because users have to improve the transmit power to meet the QoS constraints. Furthermore, we can observe that as the minimum transmission rate of users increases, the ratio of active elements decreases, while the ratio of passive elements increases. This is because when the user increases the transmit power, active elements amplifying the signal require more energy, which results in $t_{2}$ decreasing. However, the energy consumed by the passive element is not influenced by the transmit power of the user.

\section{Conclusion}

In this paper, a hybrid RIS model for energy harvesting is proposed, in which each reflecting element can flexibly operate in active, passive, and idle modes. PPO-Convex algorithm is proposed to reduce the algorithm's action space and the simulation results verified that the hybrid RIS architecture proposed in this paper is superior to other baseline schemes.

% To print the credit authorship contribution details
\printcredits

%% Loading bibliography style file
%\bibliographystyle{model1-num-names}
\bibliographystyle{elsarticle-num}

% Loading bibliography database
\bibliography{cas-refs}

% Biography
%\bio{}
% Here goes the biography details.
%\endbio

%\bio{pic1}
% Here goes the biography details.
%\endbio

\end{document}